\documentclass[11pt,fleqn]{article}

\usepackage{a4}
\usepackage{amstext}
\usepackage{amsfonts}
\usepackage{amssymb}
\usepackage{amsmath}
\usepackage{color}
\usepackage{subfig}
\usepackage{epsfig}
\usepackage{booktabs}
\usepackage{caption}
\usepackage{multirow}

\newcommand{\gtapprox}{\raisebox{-0.5ex}{$\,\stackrel{>}{\scriptstyle\sim}\,$}}
\newcommand{\ltapprox}{\raisebox{-0.5ex}{$\,\stackrel{<}{\scriptstyle\sim}\,$}}

\begin{document}

% ********************
% ********************
% ********************
% ********************
% ********************

%\begin{flushright}
%XXX XXX-XXX
%\end{flushright}

\begin{center}

{\huge {\bf Lattice investigation of the scalar mesons}

$a_0(980)$ {\bf and} $\kappa$ {\bf using four-quark operators}}

\vspace{0.5cm}

\textbf{Constantia Alexandrou}$^{1,2}$, \textbf{Jan Oliver Daldrop}$^3$, \textbf{Mattia Dalla Brida}$^4$, \\ \textbf{Mario Gravina}$^1$, \textbf{Luigi Scorzato}$^5$, \textbf{Carsten Urbach}$^3$, \textbf{Marc Wagner}$^6$

$^1$~Department of Physics, University of Cyprus, P.O.\ Box 20537, 1678 Nicosia, Cyprus

$^2$~Computation-based Science and Technology Research Center, Cyprus Institute, \\ 20 Kavafi Str., Nicosia 2121, Cyprus

$^3$~Helmholtz-Institut f{\"u}r Strahlen- und Kernphysik (Theorie) and \\ Bethe Center for Theoretical Physics, Universit{\"a}t Bonn, D-53115 Bonn, Germany

$^4$~School of Mathematics, Trinity College Dublin, Dublin 2, Ireland

$^5$~ECT$^\star$, Strada delle Tabarelle, 286, I-38123, Trento, Italy

$^6$~Goethe-Universit\"at Frankfurt am Main, Institut f\"ur Theoretische Physik, \\ Max-von-Laue-Stra{\ss}e 1, D-60438 Frankfurt am Main, Germany

\vspace{0.7cm}

\begin{picture}(0,0)%
\includegraphics{Logo.pstex}%
\end{picture}%
\setlength{\unitlength}{4144sp}%
\begingroup\makeatletter\ifx\SetFigFont\undefined%
\gdef\SetFigFont#1#2#3#4#5{%
  \reset@font\fontsize{#1}{#2pt}%
  \fontfamily{#3}\fontseries{#4}\fontshape{#5}%
  \selectfont}%
\fi\endgroup%
\begin{picture}(1620,1620)(1,-781)
\end{picture}%

\vspace{0.4cm}

%\today
December 6, 2012

\end{center}

\vspace{0.1cm}

%\textbf{XXXXX replace by a fixed date, before submitting to ArXiv XXXXX}
%\textbf{XXXXX remove ``usepackage drftcite'', before submitting to ArXiv XXXXX}
%\vspace{-0.40cm}
\begin{tabular*}{16cm}{l@{\extracolsep{\fill}}r} \hline \end{tabular*}

\vspace{-0.40cm}
\begin{center} \textbf{Abstract} \end{center}
\vspace{-0.40cm}

%\begin{abstract}
We carry out an exploratory study of the isospin one $a_0(980)$ and the isospin one-half $\kappa$ scalar mesons using $N_f=2+1+1$ Wilson twisted mass fermions
at one lattice spacing. The valence strange quark is included as an Osterwalder-Seiler  fermion  with mass tuned so that the kaon mass matches the corresponding
mass  in the unitary $N_f=2+1+1$ theory. We investigate the internal structure of these mesons by using a basis of four-quark interpolating fields. We construct diquark-diquark and molecular-type
interpolating fields and analyse the resulting correlation matrices keeping only connected contributions. For both channels, the low-lying spectrum is found to
be consistent with two-particle scattering states. Therefore, our analysis shows no evidence for an additional state that can be interpreted as either a tetraquark or a tightly-bound molecular state. 
%\end{abstract}

\begin{tabular*}{16cm}{l@{\extracolsep{\fill}}r} \hline \end{tabular*}

\thispagestyle{empty}

% ********************
% ********************
% ********************
% ********************
% ********************

\setcounter{page}{1}

\section{Introduction}

The Naive Quark Model (NQM) is -- despite its simplicity --
surprisingly successful in qualitatively describing the
experimentally observed meson and baryon spectrum. This success has led
us to think of mesons and baryons as $q \bar q$ and $qqq$ bound
states, respectively. In particular, no mesonic state incompatible
with the quantum numbers of a $q\bar{q}$ system has been confirmed,
yet. However, there are a few
exceptions~\cite{Jaffe:2004ph,Amsler:2004ps} which cannot be described
in the NQM. One prominent example is the Roper resonance, another not
less prominent one is the presence of too many scalar states (i.e with
quantum numbers $J^{PC}=0^{++}$) with mass below $2\ \mathrm{GeV}$ as
compared to the expectation from the NQM. These scalars are the
$f_0(600)$ or $\sigma$,  $f_0(980)$,  $f_0(1370)$, $f_0(1500)$ and
$f_0(1710)$ with isospin $0$, the $K_0^\ast(800)$ or $\kappa$ and
$K_0^\ast(1430)$ with isospin $1/2$, and the $a_0(980)$ and $a_0(1450)$
with isospin $1$. 

This excess of states compared to the NQM expectation suggests that
the picture of mesons as $q\bar q$ bound states is too simplistic and
it has to be complemented by other quarkonic and gluonic structures.
Consequently, it has been speculated that some of these particles are
tetraquark states, i.e.\ bound states of two quarks and two antiquarks,
or predominantly gluonic in nature. For example, according to one
favoured interpretation~\cite{Amsler:2004ps}, the states $f_0(1370),
f_0(1710)$, $K_0^*(1430)$ and  $a_0(1450)$ might indeed have dominant
$q\bar{q}$ components, as expected in the NQM but the state
$f_0(1500)$ might be, predominantly, the lightest ($0^{++}$)
glueball~\cite{Morningstar:1999rf}, and the lightest of the scalar
states might constitute a nonet with a dominant tetraquark
contribution~\cite{Jaffe:1976ig,Jaffe:1976ih,Jaffe:2004ph,Maiani:2004uc,Fariborz:2005gm,Giacosa:2006rg}.
While such an interpretation is adopted  by other authors,
e.g.\ \cite{Parganlija:2012fy}, there are also different scenarios
discussed in the literature, as for example in
\cite{Crede:2008vw,Klempt:2007cp}.

Experimentally, many of the aforementioned scalar resonances are
difficult to resolve as they have large decay widths and several decay
channels that sometimes open up only within a short energy interval.
The question whether a physical state is dominated by a $q\bar{q}$, a
tetraquark, a glueball or other hybrid wavefunctions is then typically
investigated through the analysis of its production and decay
modes. These are directly accessible in experiments and can be often
measured rather accurately. This justifies the high experimental
activity~\cite{Klempt:2007cp,PDG} in investigating the composition of
these states. It is thus crucial to develop a deeper theoretical
understanding for the internal structure of these states.

A theoretical understanding from first principles requires a non-perturbative method. 
Since Quantum Chromodynamics (QCD) is the theory of strongly coupled quarks and gluons, 
such a non-perturbative method is provided by lattice QCD. But investigating the states in
lattice QCD is also a challenging endeavour: the distinction between
scattering states, resonances and bound states is subtle on a
Euclidean lattice with finite spacetime volume. In fact, there is no continuum spectrum in a
finite spatial volume and the Hamiltonian has only discrete eigenvalues. In
order to disentangle these different physical phenomena it is
necessary to study the volume dependence of the discrete eigenvalues
of the
Hamiltonian~\cite{Luscher:1986pf,Luscher:1990ux,Luscher:1991cf,Wiese:1988qy}.
In particular, the coefficients of the large volume expansion of the
discrete eigenvalues are related to the phase shifts of the scattering
process.  Moreover, as the volume increases, the eventual resonances
produce "avoided level crossings" of eigenvalues~\footnote{Note,
  however, that also the threshold may display the same
  phenomenon~\cite{Doring:2012eu}.}. 

This method requires the extraction of more than the ground state in a
channel with given quantum numbers. These excited states  
are increasingly difficult to extract with sufficient
precision, even though the field has recently seen tremendous progress
in the methodology. An additional complication
is the appearance of fermionic disconnected contributions, which are
notoriously noisy. Therefore, the available lattice results on scalar
mesons and possibly existing tetraquark states are still limited (cf.\
e.g.\
\cite{Prelovsek:2008rf,Jansen:2009hr,Prelovsek:2010kg,Alexandrou:2004ak,Okiharu:2004ve,Wagner:2011ev,Bicudo:2012qt,Kalinowski:2012re}).
Certainly more and in particular independent investigations are needed
to gain a better understanding of these scalar states.

In this paper we perform an exploratory study of the $a_0(980)$ and
the $\kappa$ using Wilson twisted mass fermions. It is the first study
of this kind with $N_f=2+1+1$ dynamical quark flavours, using gauge
configurations provided by the European Twisted Mass (ETM)
collaboration~\cite{Baron:2008xa,Baron:2009zq,Baron:2010bv,Baron:2011sf}.
Note that in particular the dynamical strange quark might be important
for studying scalar resonances.
In this exploratory investigation we address the question, whether or
not these states 
could be consistent with a tetraquark or a molecule interpretation. We
focus on the precise computation of correlation functions of operators
with quantum numbers of the $a_0(980)$ and $\kappa$ mesons using four
quark interpolating fields ignoring fermionic disconnected
contributions. The latter implies that there is no mixing among four
quark, two-quark and gluonic states. Apart from obvious
technical advantages and having a testbed of our method, there is another important reason for working in this approximation: in \cite{Prelovsek:2010kg}
bound states close to threshold in the $I=0$ and the $I=1/2$ channels
have been found in the same approximation. These bound states, found
in addition to the expected scattering states, were interpreted as a
possible indication for a tetraquark nature of the corresponding
states. In our study, performed with a similar operator basis, but a different lattice discretisation, we do not observe such a bound
state in the $I=1/2$ channel. Moreover, we also do not observe it
in the $I=1$ channel, which was not considered in
\cite{Prelovsek:2010kg}. Note that parts of this work have recently
been presented in a conference proceeding~\cite{Daldrop:2012sr}.

The reason for focussing on the $a_0$ and the $\kappa$ are the
following: the $a_0$ has isospin $I = 1$, i.e.\ when choosing $I_z =
\pm 1$ only a single disconnected 
contribution is ignored. The $\kappa$ meson, on the other hand,
mixes only with the $K + \pi$ channel. 

The paper is organised as follows: in section \ref{SEC338} we
introduce the lattice formulation followed by a discussion of the
operator basis in section~\ref{SEC487}; the results of our study are
discussed in section~\ref{SEC448} and we conclude in the last section.

% ********************
% ********************
% ********************
% ********************
% ********************

\section{\label{SEC338}Lattice setup}

% **********
% **********
% **********

\subsection{\label{SEC693}Lattice actions}

This work is based on gauge link configurations generated by the ETM collaboration~\cite{Baron:2008xa,Baron:2009zq,Baron:2010bv,Baron:2011sf} with the Iwasaki gauge action~\cite{Iwasaki:1985we} and $N_f = 2+1+1$ flavours of twisted mass quarks.

The light degenerate $(u,d)$ quark doublet is described by the standard Wilson twisted mass action \cite{Frezzotti:2000nk},
\begin{eqnarray}
\label{EQN773} S_{\textrm{light}}[\chi^{(l)},\bar{\chi}^{(l)},U] \ \ = \ \ a^4 \sum_x \bar{\chi}^{(l)}(x) \Big(D_\textrm{W}(m_0) + i \mu \gamma_5 \tau_3\Big) \chi^{(l)}(x) ,
\end{eqnarray}
while for the heavy $(c,s)$ sea quark doublet the twisted mass formulation for non-degenerate quarks of \cite{Frezzotti:2003xj} has been used,
\begin{eqnarray}
\label{EQN004} S_{\textrm{heavy}}[\chi^{(h)},\bar{\chi}^{(h)},U] \ \ = \ \ a^4 \sum_x \bar{\chi}^{(h)}(x) \Big(D_\textrm{W}(m_0) + i \mu_\sigma \gamma_5 \tau_1 + \tau_3 \mu_\delta\Big) \chi^{(h)}(x) .
\end{eqnarray}
In both cases $D_\mathrm{W}$ denotes the standard Wilson Dirac operator,
\begin{eqnarray}
D_\mathrm{W}(m_0) \ \ = \ \ \frac{1}{2} \Big(\gamma_\mu \Big(\nabla_\mu + \nabla^\ast_\mu\Big) - a \nabla^\ast_\mu \nabla_\mu\Big) + m_0 ,
\end{eqnarray}
while $\chi^{(l)} = (\chi^{(u)},\chi^{(d)})$ and $\chi^{(h)} = (\chi^{(c)},\chi^{(s)})$ are the quark fields in the so-called twisted basis. For reasons explained in \cite{Chiarappa:2006ae} the same value of the standard quark mass parameter $m_0$ has been used in both sectors.

When tuning the theory to maximal twist, automatic $\mathcal{O}(a)$ improvement for physical quantities applies~\cite{Frezzotti:2003xj,Frezzotti:2003ni}. This tuning has been done by adjusting $m_0$ such that the PCAC quark mass in the light quark sector vanishes (cf.\ \cite{Baron:2010bv} for details).

At maximal twist in a massless quark renormalisation scheme the renormalised quark masses are related to the bare parameters $\mu_\sigma$ and $\mu_\delta$ by
\begin{eqnarray}
m^R_{s} \ \ = \ \ Z_P^{-1} \bigg(\mu_\sigma - \frac{Z_P}{Z_S} \mu_\delta\bigg) \qquad , \qquad m^R_{c} \ \ = \ \ Z_P^{-1} \bigg(\mu_\sigma + \frac{Z_P}{Z_S} \mu_\delta\bigg)
\end{eqnarray}
\cite{Frezzotti:2003xj}, where $Z_P$ and $Z_S$ are the renormalisation constants of the non-singlet pseudoscalar and scalar densities. In our simulations the values of $\mu_\sigma$ and $\mu_\delta$ have been adjusted by requiring that the resulting lattice kaon and $D$ meson masses approximately assume their physical values~\cite{Baron:2010bv,Baron:2010th,Baron:2010vp}.

For the computation of observables we use a twisted mass discretisation for valence $s$ quarks, which is different from the sea $s$ quarks (\ref{EQN004}). It is given by (\ref{EQN773}) with $\chi^{(l)} \rightarrow \chi^{(s)} = (\chi^{(s^+)} , \chi^{(s^-)})$ and $\mu_l \rightarrow \mu_s$. We do this, to avoid the problem of mixing between $s$ and $c$ quarks, which is discussed in detail in \cite{Baron:2010th,Baron:2010vp}. Note that there are two possibilities to realize a valence $s$ quark, $\chi^{(s^+)}$ and $\chi^{(s^-)}$, which differ in the sign of the twisted mass term, $\pm i \mu_s \gamma_5$. Strategies and consequences of choosing $s^+$ or $s^-$ are discussed in detail in sections~\ref{SEC648} and \ref{SEC649}. The bare strange quark mass $\mu_s$ has been chosen such that kaon masses computed within this mixed action setup with flavour structure $\bar{s}^+ d$ and $\bar{s}^- u$ (which are degenerate and known to have less discretisation errors than their $\bar{s}^+ u$ and $\bar{s}^- d$ counterparts \cite{Sharpe:2004ny,Frezzotti:2005gi,Blossier:2007vv}) agree with kaon masses computed in the unitary setup \cite{Baron:2010th,Baron:2010vp}, i.e.\ using (\ref{EQN004}) also for valence $s$ quarks.

In this work we consider six gauge link ensembles with simulation parameters given in  Table~\ref{TAB001}. They differ in the space-time volume $(L/a)^3 \times T/a$ and in the light $u/d$ quark mass $\mu_l$. The lattice spacing $a \approx 0.086 \, \textrm{fm}$ is the same for all ensembles. More details regarding these ensembles can be found in \cite{Baron:2010bv}.

\begin{table}[htb]
\centering
\begin{tabular}{cccccccccc}
\toprule
\multirow{2}*{Ensemble} & \multirow{2}*{$\beta$} & \multirow{2}*{$(L/a)^3 \times T/a$} & \multirow{2}*{$\mu_l$} & \multirow{2}*{$\mu_\sigma$} & \multirow{2}*{$\mu_\delta$} & \multirow{2}*{$\mu_s$} & $a$ & $m_\textrm{PS}$   & \# of \\
                        &                        &                                     &                         &                         &                              &                              & $(\text{fm})$  & $(\text{MeV})$  & configs \\
\midrule
A30.32 & $1.90$ & $32^3 \times 64$ & $0.0030$ & $0.150$ & $0.190$ & $0.02280$ & $0.086$ & $284$ & $672$ \\
A40.32 &	& $32^3 \times 64$ & $0.0040$ &         &         & $0.02322$ & 	  & $324$ & $200$ \\
A40.24 &	& $24^3 \times 48$ & $0.0040$ &         &         & $0.02300$ & 	  & $332$ & $1259$ \\
A40.20 &	& $20^3 \times 48$ & $0.0040$ &         &         & $0.02308$ &	  & $341$ & $500$ \\
A50.32 &	& $32^3 \times 64$ & $0.0050$ &         &         & $0.02336$ &	  & $362$ & $431$ \\
A80.24 &	& $24^3 \times 48$ & $0.0080$ &         &         & $0.02328$ &	  & $455$ & $1225$ \\
\bottomrule
\end{tabular}
\caption{\label{TAB001}Gauge link ensembles considered in this paper. The notation follows \cite{Baron:2010bv}.} 
\end{table}

The discussion of meson and four-quark creation operators (cf.\ section~\ref{SEC487}) and their quantum numbers is more convenient with quark fields in the ``physical basis'', $(u,d)$ and $(s^+,s^-)$. This physical basis is related to the ``twisted basis'' $(\chi^{(u)},\chi^{(d)})$ and $(\chi^{(s^+)},\chi^{(s^-)})$ introduced in (\ref{EQN773}) and (\ref{EQN004}) according to
\begin{eqnarray}
\label{EQN001} \bigg(\begin{array}{c} u \\ d \end{array}\bigg) \ \ = \ \ e^{i \gamma_5 \tau_3 \omega / 2} \bigg(\begin{array}{c} \chi^{(u)} \\ \chi^{(d)} \end{array}\bigg) \quad , \quad \bigg(\begin{array}{c} s^+ \\ s^- \end{array}\bigg) \ \ = \ \ e^{i \gamma_5 \tau_3 \omega / 2} \bigg(\begin{array}{c} \chi^{(s^+)} \\ \chi^{(s^-)} \end{array}\bigg) ,
\end{eqnarray}
where $\omega$ is the twist angle, which we have tuned to maximal twist, i.e.\ $\omega = \pi/2$.

\begin{figure}[hbtp]
\centering
\subfloat[][\emph{Connected contribution}.]
{\includegraphics[width=.40\columnwidth]{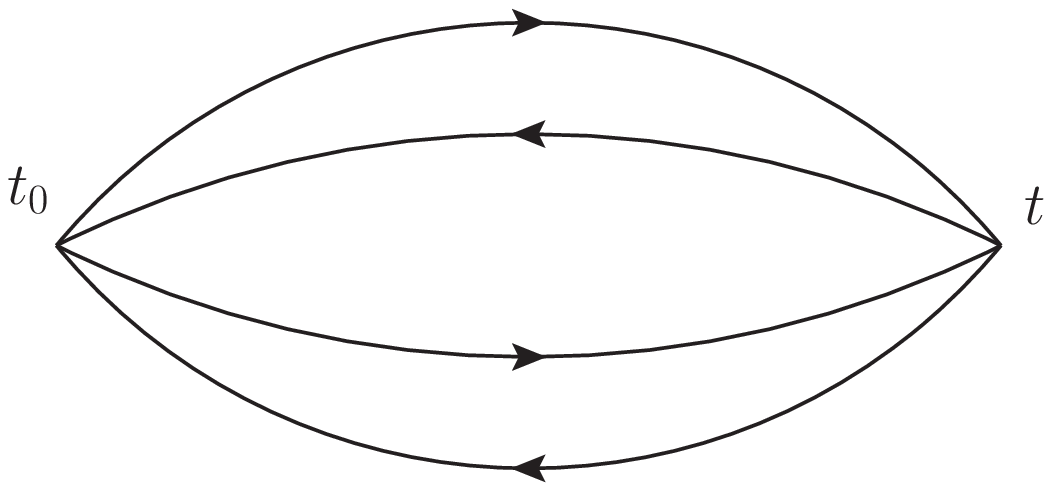}} \quad
\subfloat[][\emph{Singly disconnected contribution}.]
{\includegraphics[width=.40\columnwidth]{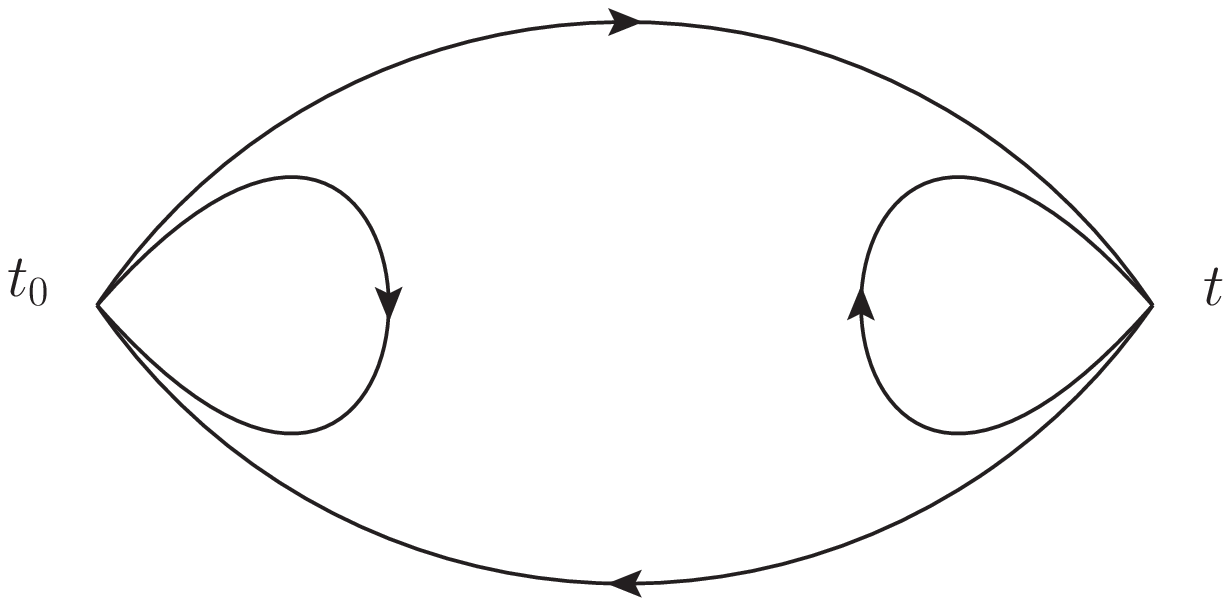}} \\
\subfloat[][\emph{Doubly disconnected contribution}.]
{\includegraphics[width=.40\columnwidth]{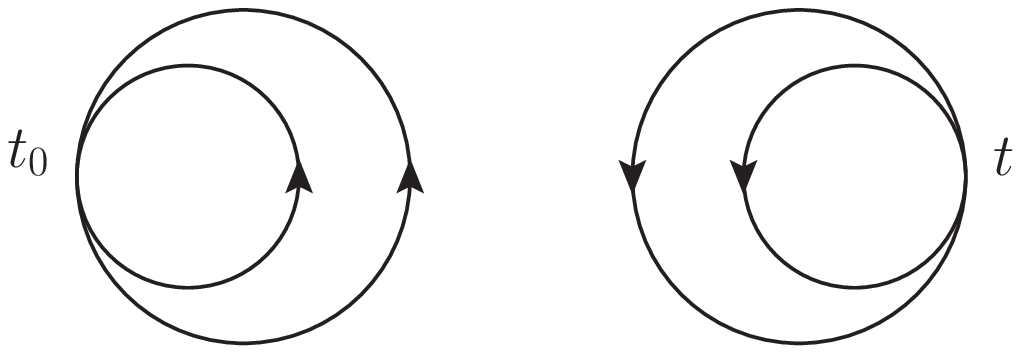}} 
\caption{Relevant contributions to a four-quark operator two point function.}
\label{fig:diagrams}
\end{figure}

When computing temporal correlation functions $\langle \mathcal{O}_j^\dagger(t_2) \mathcal{O}_k(t_1) \rangle$, where $\mathcal{O}_j$ and $\mathcal{O}_k$ are e.g.\ meson or four-quark creation operators, 
we only consider quark propagators connecting time $t_1$ and $t_2$, but ignore propagation of quarks within the same timeslice, 
e.g.\ from $t_1$ to $t_1$. For mesons this amounts to neglecting so-called disconnected diagrams. 
For four-quark operators e.g.\ of tetraquark or two-meson type both singly and doubly disconnected contributions (cf. (b) and (c) of Figure \ref{fig:diagrams}) are omitted. 
Consequences of not considering disconnected diagrams are discussed in the following sections.

Finally it should be mentioned that at finite lattice spacing isospin
and parity are not good quantum numbers in twisted mass lattice QCD. These
symmetries are broken by terms proportional to
$\mathcal{O}(a)$. Consequently, when doing spectroscopy, one has to
take mixing with states of opposite parity and different isospin into
account. Of course, in the continuum limit these symmetries are
restored and QCD is recovered. Mixing in the context of $a_0(980)$ and
$\kappa$ is discussed in detail in sections \ref{SEC328} and
\ref{SEC329}.

% **********
% **********
% **********

\subsection{\label{SEC537}The pseudoscalar meson spectrum}

Since a pair of pseudoscalar mesons is rather light and can have the
same quantum numbers as the scalar mesons $a_0(980)$ and $\kappa$,
these are the most relevant scattering states to consider. Their masses are approximately equal to the sum of the two masses of the corresponding individual mesons. Therefore, a precise and comprehensive knowledge of the meson spectrum is important for our analysis.

As we ignore 
disconnected contributions, the $\eta \equiv
\bar{u} u + \bar{d} d$ meson and the neutral $\pi \equiv \bar{u} u -
\bar{d} d$ become degenerate and there is an $\eta/\eta'$-like meson
with valence quark structure $\bar{s} s$, but no light $\bar{u} u$ or
$\bar{d} d$ valence quarks, which we denote by $\eta_s$. 

Another particularity stems from the valence action used for
the strange quarks discussed above. The kaon and the $\eta_s$ can be constructed using $s^+$ and/or $s^-$ strange quarks, resulting in different values for the meson masses at finite
value of the lattice spacing. Similarly, in Wilson twisted mass lattice
QCD the neutral (connected-only) and charged pion mass values differ.

All meson masses relevant for our investigation are collected
in Table~\ref{TAB011}.

\begin{table}[htb]
\centering
\begin{tabular}{ccccccc}
\toprule
\multirow{2}*{ensemble} & \multirow{2}*{$m_{\pi (\bar{u} d, \bar{d} u)}$} &
\multirow{2}*{$m_{\pi (\bar{u} u - \bar{d} d)}$} & \multirow{2}*{$m_{K (\bar{s}^+
d, \bar{s}^- u)}$} & \multirow{2}*{$m_{K (\bar{s}^+ u, \bar{s}^- d)}$} &
\multirow{2}*{$m_{\eta_s (\bar{s}^+ s^-)}$} & \multirow{2}*{$m_{\eta_s (\bar{s}^+
s^+, \bar{s}^- s^-)}$}  \\
                        &                        &
&                         &                         &
                            &                             \\
\midrule
A30.32 & 284(1) & 494(6) & 576(7) & 704(2) &  & 876(1) \\
A40.20 & 341(2) &  & 599(3) & & 774(2) & \\
% A40.20 & 341(2) &  & 599(3) & 715(6) & 774(2) & 873(6) \\
%
A40.24 & 332(1) & 530(7) & 593(1) & 723(2) &  & 882(1) \\
A40.32 & 324(7) &  & 588(5) & & 779(2) & \\
% A40.32 & 324(7) &  & 588(5) & 728(6) & 779(2) & 884(2) \\
%
A50.32 & 362(7) &  & 601(9) &  & 783(2) &  \\
A80.24 & 455(1) & 625(3) & 635(1) & 753(1) &  & 885(1) \\
\bottomrule
\end{tabular}
\caption{\label{TAB011}The pseudoscalar meson spectrum in MeV with disconnected diagrams neglected. Omitted mass values are not needed in the context of the tetraquark study presented in this paper.}
\end{table}

% ********************
% ********************
% ********************
% ********************
% ********************

\section{\label{SEC487}Creation operators and analysis details}

In this study we exclusively consider creation operators with four quarks (two quarks and two antiquarks). The structure of our four-quark operators is oriented at phenomenological expectations and ranges from four-quark bound states (molecules formed by two mesons and bound diquark-antidiquark pairs) to two essentially non-interacting mesons (two-particle operators).

Of course, standard quark-antiquark operators, e.g.\ $\bar{d} u$ for $a_0(980)$ and $\bar{s} u$ for $\kappa$, would also be of interest. However, since we neglect disconnected diagrams (cf.\ section~\ref{SEC693}), such two-quark operators do not generate overlap to trial states created by four-quark operators. Consequently, in our setup four-quark operators and quark-antiquark operators probe different sectors, which is, why we do not consider the latter in the following. In a subsequent improved study we plan to include disconnected diagrams and to combine two- and four-quark operators in a single correlation matrix.

Information regarding the used four-quark operators is summarised in Table~\ref{TAB569}. The operators will be discussed in more detail below.

\begin{table}[htb]
\centering
\begin{tabular}{cccccc}
\toprule
\multirow{2}*{ensemble} & quark    & gauge    & \multirow{2}*{type} & \multicolumn{2}{c}{Dirac structure} \\ 
\cmidrule(lr){5-6}                        & smearing & smearing &                    & $a_0(980)$ & $\kappa$ \\
\midrule
A30.32   & \multirow{4}*{Gaussian}   &   \multirow{4}*{APE}    &  $K \bar{K} \textrm{ molecule}$    & $\gamma_5,\gamma_\mu,\gamma_\mu\gamma_5$ & $-$  \\
\multirow{2}*{A40.24}     &                &                       & $\eta_s \pi \textrm{ molecule}$ & $\gamma_5$  &  $-$   \\
         &                &                       &  $K \pi \textrm{ molecule}$    & $-$   &  $\gamma_5,\gamma_\mu,\gamma_\mu\gamma_5$ \\
A80.24   &                &                       &  diquark               & $C\gamma_5,C$  &  $C\gamma_5,C$   \\
\midrule
\multirow{8}*{A40.20}   & \multirow{4}*{no}    &  \multirow{4}*{no}   & $K \bar{K} \textrm{ molecule}$    & $\gamma_5$   & $-$ \\
                        &                &                       & diquark               & $C\gamma_5$  & $-$   \\
                        &                &                       & $K + \bar{K} \textrm{ two-particle}$     & $\gamma_5$  & $-$   \\		
                        &                &                       & $\eta_s + \pi \textrm{ two-particle}$  & $\gamma_5$  & $-$   \\		
\cmidrule(lr){2-6}      & \multirow{3}*{Gaussian} & \multirow{3}*{APE}  & $K \bar{K} \textrm{ molecule}$    & $\gamma_5,\gamma_\mu$   & $-$ \\
						&				 &						 & $\eta_s \pi \textrm{ molecule}$ & $\gamma_5$  & $-$   \\		
						&				 &						 & diquark               & $C\gamma_5,C$  & $-$   \\
\midrule
\multirow{3}*{A40.32}   & \multirow{3}*{Gaussian} & \multirow{3}*{APE}  & $K \bar{K} \textrm{ molecule}$    & $\gamma_5,\gamma_\mu$   & $-$ \\
                        &     			 &						 & $\eta_s \pi \textrm{ molecule}$ & $\gamma_5$  & $-$   \\		
                        &     			 &						 & diquark               & $C\gamma_5,C$  & $-$   \\
\midrule
\multirow{2}*{A50.32 }  & \multirow{2}*{Gaussian} & \multirow{2}*{APE}  & $K \bar{K} \textrm{ molecule}$    & $\gamma_5,\gamma_\mu$   & $-$ \\
                        &     			 &						 & diquark               & $C\gamma_5,C$  & $-$   \\
\bottomrule
\end{tabular}
\caption{\label{TAB569}Four-quark creation operators.}
\end{table}

% **********
% **********
% **********

\subsection{\label{SEC483}Creation operators, $a_0(980)$ sector (quantum numbers $I(J^P) = 1(0^+)$)}

The expected low-lying spectrum in the $a_0(980)$ sector ($\approx 1000 \, \textrm{MeV}$) is the following:
\begin{itemize}
\item A two-particle $\eta + \pi$ and a two-particle $\eta' + \pi$ state.
\begin{itemize}
\item In nature:
\begin{itemize}
\item Mass $m(\eta + \pi) \approx 548 \, \textrm{MeV} + 140 \, \textrm{MeV} = 688 \, \textrm{MeV}$ \cite{PDG}.

\item Mass $m(\eta' + \pi) \approx 958 \, \textrm{MeV} + 140 \, \textrm{MeV} = 1098 \, \textrm{MeV}$ \cite{PDG}.
\end{itemize}

\item In our lattice setup:
\begin{itemize}
\item Due to neglect of disconnected diagrams $\eta$ has flavour structure $\bar{u} u + \bar{d} d$ and is degenerate with the neutral pion (cf.\ section~\ref{SEC537}); the $\eta + \pi$ state is orthogonal to any trial state obtained by using an operator containing $s$ quarks, i.e.\ can be ignored in the following.

\item Due to neglect of disconnected diagrams $\eta'$ becomes $\eta_s$ (cf.\ section~\ref{SEC537}); masses $m(\eta_s + \pi) \approx m(\eta_s) + m(\pi)$ depend on the gauge link ensemble and can be read off from Table~\ref{TAB011}.
\end{itemize}
\end{itemize}

\item A two-particle $K + \bar{K}$ state.
\begin{itemize}
\item In nature: mass $m(K + \bar{K}) \approx 2 \times 496 \, \textrm{MeV} = 992 \, \textrm{MeV}$ \cite{PDG}.

\item In our lattice setup: masses $m(K + \bar{K}) \approx 2 m(K)$ depend on the gauge link ensemble and can be read off from Table~\ref{TAB011}.
\end{itemize}

\item Possibly a bound $a_0(980)$ state (might be of quark-antiquark, molecule or diquark-antidi\-quark type), mass $m(a_0(980)) = 980 \pm 20 \, \textrm{MeV}$ \cite{PDG}.
\end{itemize}

To be able to resolve these low-lying states, we consider the following operators:
\begin{itemize}
\item Molecule type operators:
\begin{eqnarray}
\label{EQN002} & & \hspace{-0.7cm} \mathcal{O}_{a_0(980)}^{K \bar{K} \textrm{ molecule}} \ \ = \ \ \sum_\mathbf{x} \Big(\bar{s}(\mathbf{x}) \gamma_5 u(\mathbf{x})\Big) \Big(\bar{d}(\mathbf{x}) \gamma_5 s(\mathbf{x})\Big) \\
\label{EQN002_} & & \hspace{-0.7cm} \mathcal{O}_{a_0(980)}^{\eta_s \pi \textrm{ molecule}} \ \ = \ \ \sum_\mathbf{x} \Big(\bar{s}(\mathbf{x}) \gamma_5 s(\mathbf{x})\Big) \Big(\bar{d}(\mathbf{x}) \gamma_5 u(\mathbf{x})\Big) ;
\end{eqnarray}
since pseudoscalar mesons (mesons with spin structure $\gamma_5$) are the lightest mesons in a given flavour sector, one expects possible molecular bound states to be of pseudoscalar-pseudoscalar type. We also consider molecule type operators with $\gamma_5$ replaced by $\gamma_j$ and by $\gamma_j \gamma_5$. These operators enlarge our correlation matrices and allow us to study also excited states, in particular two-particle states with relative momentum (cf.\ section~\ref{SEC448}).

\item Diquark type operator:
\begin{eqnarray}
\label{EQN524} \mathcal{O}_{a_0(980)}^{\textrm{diquark}} \ \ = \ \ \sum_\mathbf{x} \Big(\epsilon^{a b c} \bar{s}^b(\mathbf{x}) C \gamma_5 \bar{d}^{c,T}(\mathbf{x})\Big) \Big(\epsilon^{a d e} u^{d,T}(\mathbf{x}) C \gamma_5 s^e(\mathbf{x})\Big) ;
\end{eqnarray}
diquarks with spin structure $\gamma_5$ are known to be the lightest \cite{Jaffe:2004ph,Alexandrou:2006cq,Wagner:2011fs}, which is, why we use $\gamma_5$ in this operator. We also consider diquark type operators with $\gamma_5$ replaced by $1$. As before, the main reason is to enlarge our correlation matrices allowing us to study also excited states.

\item Two-particle type operators:
\begin{eqnarray}
\label{EQN093} & & \hspace{-0.7cm} \mathcal{O}_{a_0(980)}^{K + \bar{K} \textrm{ two-particle}} \ \ = \ \ \bigg(\sum_\mathbf{x} \bar{s}(\mathbf{x}) \gamma_5 u(\mathbf{x})\bigg) \bigg(\sum_\mathbf{y} \bar{d}(\mathbf{y}) \gamma_5 s(\mathbf{y})\bigg) \\
\label{EQN003} & & \hspace{-0.7cm} \mathcal{O}_{a_0(980)}^{\eta_s + \pi \textrm{ two-particle}} \ \ = \ \ \bigg(\sum_\mathbf{x} \bar{s}(\mathbf{x}) \gamma_5 s(\mathbf{x})\bigg) \bigg(\sum_\mathbf{y} \bar{d}(\mathbf{y}) \gamma_5 u(\mathbf{y})\bigg) ;
\end{eqnarray}
these operators resemble states with two non-interacting mesons and, therefore, should be particularly suited to resolve two-particle $K + \bar{K}$ and $\eta_s + \pi$ states. Note that terms with $\mathbf{x} = \mathbf{y}$ in (\ref{EQN093}) and (\ref{EQN003}) also appear in the molecule operators (\ref{EQN002}) and (\ref{EQN002_}), which is, why two-particle $K + \bar{K}$ and $\eta_s + \pi$ states can also be resolved, even though only molecule and diquark operators are used. However, the generated overlap to two-particle states is significantly larger, when two-particle operators are applied, which in turn results in a signal of better statistical quality (cf.\ the numerical results in section~\ref{SEC129}).
\end{itemize}

% *****

\subsubsection{\label{SEC328}Mixing due to twisted mass symmetry breaking}

In twisted mass lattice QCD both parity $P$ and isospin $I$ are broken by $\mathcal{O}(a)$. Consequently, one has to take mixing with states of opposite parity and different isospin into account. When there is mixing with rather light states (lighter than those one is interested in), problems arise: correlators are slightly contaminated by weakly decaying exponentials, which become dominant at large temporal separations, at which masses are usually determined.

Before we discuss mixing due to twisted mass symmetry breaking, it is important, to understand the effects arising by neglecting disconnected diagrams in more detail. In such a setup the valence quark flavour structure is conserved, i.e.\ quark-antiquark pairs can neither be created nor annihilated. For $a_0(980)$ this implies that any state that mixes with the trial states created by our operators (cf.\ (\ref{EQN002}) to (\ref{EQN003})) must have valence quark flavour structure $u \bar{d} s \bar{s}$.

This observation is particularly important, when discussing parity mixing, because at first glance there seem to be states of opposite (negative) parity, which are light, namely pions $I(J^P) = 1(0^-)$. However, since pions only have a $u$ and a $\bar{d}$ valence quark, but no $s \bar{s}$ pair, they are orthogonal to any state we probe with our four-quark operators. On the other hand $u \bar{d} s \bar{s}$ four-quark states with negative parity are expected to be rather heavy, e.g.\ could be a pseudoscalar meson and a scalar meson like $K + \kappa$.

Since the $z$-component of isospin $I_z$ is a quantum number, and since we study the $I_z = +1$ sector, isospin mixing can only take place with $I \geq 2$ states. In principle there could be mixing with rather light $I = 2$ $\pi + \pi$ states, but as mentioned above this is prevented by neglecting disconnected diagrams, which enforce valence flavour structure $u \bar{d} s \bar{s}$, i.e.\ $I = 1$ and $I_z = +1$.

To summarise, for the $a_0(980)$ sector $I(J^P) = 1(0^+)$ mixing due to twisted mass symmetry breaking is not expected to cause any problems. This is confirmed by our numerical results (no additional unexpected states are observed, the effective mass plateaux quality is good and does not seem to be contaminated by mixing; cf.\ section~\ref{SEC448}).

% *****

\subsubsection{\label{SEC648}Different twisted mass realizations of the $s$ quark}

In our mixed action setup the $s$ quark can be realized with either a twisted mass term $+i \mu_s \gamma_5$ or $-i \mu_s \gamma_5$ denoted by $s^+$ and $s^-$, respectively (cf.\ also section~\ref{SEC693}). Consequently, the $s \bar{s}$ pair appearing in our creation operators can be $s^+ \bar{s}^+$, $s^- \bar{s}^-$, $s^+ \bar{s}^-$ or $s^- \bar{s}^+$. In the continuum limit all four choices yield identical results. At finite lattice spacing, however, results are different due to discretisation errors.

For mesons, it is known that using a quark and an antiquark with different twisted mass signs significantly reduces discretisation errors \cite{Sharpe:2004ny,Frezzotti:2005gi,Blossier:2007vv}. With this in mind $s^+ \bar{s}^-$ should be the optimal choice for the operators (\ref{EQN002}), (\ref{EQN093}) and (\ref{EQN003}).

It is not clear, whether this mixed realization is optimal also for diquarks (operator (\ref{EQN524})). For this reason, we also used $s^+ \bar{s}^+$ (or $s^- \bar{s}^-$, which yields exactly the same result). Another advantage is the possibility to also compute disconnected diagrams (which we plan to do in the near future), which is not possible, when using $s^+ \bar{s}^-$.

Performing computations both with $s^+ \bar{s}^+$ as well as with $s^+ \bar{s}^-$ is not only a valuable cross check of numerical results, but also provides a first estimate of the magnitude of discretisation errors at our current value of the lattice spacing. In section~\ref{SEC448} numerical results are presented and discussed in this context.

% **********
% **********
% **********

\subsection{\label{SEC484}Creation operators, $\kappa$ sector (quantum numbers $I(J^P) = 1/2(0^+)$)}

The expected low-lying spectrum in the $\kappa$ sector ($\approx 700 \, \textrm{MeV}$) is the following:
\begin{itemize}
\item A two-particle $K + \pi$ state.
\begin{itemize}
\item In nature: mass $m(K + \pi) \approx 496 \, \textrm{MeV} + 140 \, \textrm{MeV} = 636 \, \textrm{MeV}$ \cite{PDG}.

\item In our lattice setup: masses $m(K + \pi) \approx m(K) + m(\pi)$ depend on the gauge link ensemble and can be read off from Table~\ref{TAB011}.
\end{itemize}

\item Possibly a bound $\kappa$ state (might be of quark-antiquark, molecule or diquark-antidi\-quark type), mass
  $m = 682 \pm 29 \, \textrm{MeV}$ \cite{PDG}. Such a state has been observed in the lattice study reported in
  \cite{Prelovsek:2010kg}, but not in the one in \cite{Lang:2012sv}.  Note that disconnected contributions are
  neglected in our calculations, like in \cite{Prelovsek:2010kg}.
\end{itemize}

To be able to resolve these low-lying states, we proceed similar as for $a_0(980)$ and consider the following operators:
\begin{itemize}
\item Molecule type operator:
\begin{eqnarray}
\nonumber & & \hspace{-0.7cm} \mathcal{O}_\kappa^{K \pi \textrm{ molecule}} \ \ = \\
\nonumber & & = \ \ \sum_\mathbf{x} \bigg(\Big(\bar{s}(\mathbf{x}) \gamma_5 u(\mathbf{x})\Big) \Big(\bar{u}(\mathbf{x}) \gamma_5 u(\mathbf{x})\Big) + \Big(\bar{s}(\mathbf{x}) \gamma_5 d(\mathbf{x})\Big) \Big(\bar{d}(\mathbf{x}) \gamma_5 u(\mathbf{x})\Big) \\
\label{EQN522} & & \hspace{0.675cm} + \Big(\bar{s}(\mathbf{x}) \gamma_5 s(\mathbf{x})\Big) \Big(\bar{s}(\mathbf{x}) \gamma_5 u(\mathbf{x})\Big)\bigg) ;
\end{eqnarray}
to be able to check and compare with results of a recent similar lattice tetraquark study of $\kappa$ \cite{Prelovsek:2010kg}, we also consider molecule type operators with $\gamma_5$ replaced by $\gamma_j$ and by $\gamma_j \gamma_5$. Such a structure corresponds to a bound state of a pair of vector mesons ($\gamma_j$) and pair of axial vector mesons ($\gamma_j \gamma_5$), which are significantly heavier than pseudoscalar mesons ($\gamma_5$). Therefore, we do not expect them to be very helpful to resolve low lying states, which is confirmed by our numerical results (cf.\ section~\ref{SEC001}). They are, however, useful for the extraction of excited states.

\item Diquark type operator:
\begin{eqnarray}
\label{EQN523} \mathcal{O}_\kappa^{\textrm{diquark}} \ \ = \ \ \sum_\mathbf{x} \Big(\epsilon^{a b c} \bar{s}^b(\mathbf{x}) C \gamma_5 \bar{d}^{c,T}(\mathbf{x})\Big) \Big(\epsilon^{a d e} d^{d,T}(\mathbf{x}) C \gamma_5 u^e(\mathbf{x})\Big) ;
\end{eqnarray}
note that $\gamma_5$ diquark flavour combinations $[\bar{s} \bar{u}] [u u]$ and $[\bar{s} \bar{s}] [s u]$ do not exist, due to the Grassmann property of the quark fields, i.e.\ $[u u] = [\bar{s} \bar{s}] = 0$. Hence, in contrast to the molecule operator (\ref{EQN522}) there is no sum over light quark flavours in the diquark operator (\ref{EQN523}); as before, we also consider diquark type operators with $\gamma_5$ replaced by $1$; since a diquark with spin structure $1$ is known to be heavier than a diquark with spin structure $\gamma_5$ \cite{Alexandrou:2006cq,Wagner:2011fs}, we mainly use it to resolve excited states.
\end{itemize}

Both our numerical results for $a_0(980)$ and the above mentioned lattice study of $\kappa$ \cite{Prelovsek:2010kg} indicate that two-particle scattering states can be resolved with four-quark operators, where all quarks are located at the same point (e.g.\ operators (\ref{EQN522}) and (\ref{EQN523})). Therefore, in contrast to our study of $a_0(980)$ we do not consider operators of two-particle type. This allows to save a significant amount of computer time, because two-particle operators require different inversions of the Dirac operator (timeslice propagators instead of point propagators; cf.\ section~\ref{SEC229}).

% *****

\subsubsection{\label{SEC329}Mixing due to twisted mass symmetry breaking}

In contrast to the $a_0(980)$ sector, mixing introduces additional low-lying states in the $\kappa$ sector, which have to be understood and resolved. These additional states have their origin in two-particle $K + \pi$ states with $I = 3/2$ (an $I = 1/2$ kaon and an $I = 1$ pion can either form an $I = 1/2$ or $I = 3/2$ state).

In QCD, where isospin is conserved, these states are degenerate. One can linearly combine the degenerate $I_z \in \{ -1/2 , +1/2 \}$ kaons and $I_z = \{ -1, 0, +1 \}$ pions with appropriate Clebsch Gordan coefficients to form states with defined isospin $I = 1/2$ and $I = 3/2$. Note, however, that any other linear combination is still an eigenstate of the QCD Hamiltonian, i.e.\ when discussing eigenstates of the Hamiltonian defined isospin is an option, but not a necessity.

In twisted mass lattice QCD isospin is broken by $\mathcal{O}(a)$, i.e.\ $u$ and $d$ as well as $s^+$ and $s^-$ quarks are treated differently. The important consequence in the context of this discussion is that the $I_z = -1/2$ kaon $\bar{s}^+ d$ is lighter than the $I_z = +1/2$ kaon $\bar{s}^+ u$. Similarly there is a splitting of pion masses, where the $I_z = \pm 1$ pions ($\bar{d} u$ and $\bar{u} d$) are lighter than their $I_z = 0$ counterparts ($\bar{u} u - \bar{d} d$). While in QCD any linear combination of these kaons and pions is an eigenstate of the Hamiltonian, this splitting determines specific linear combinations, which are eigenstates in twisted mass lattice QCD: there is a $(K + \pi) \equiv ( \bar{s}^+ u  + (\bar{u} u - \bar{d} d))$ state and a $(K + \pi) \equiv (\bar{s}^+ d + \bar{d} u)$ state; the two mesons in the first state are heavier than the two mesons in the second state (cf.\ Table~\ref{TAB011}). Note that both combinations have $I = 1/2$ and $I = 3/2$ contributions of the same order 
of magnitude, i.e.\ are not even close to isospin eigenstates. Thus, when determining the low lying spectrum, one needs to resolve $I = 1/2$ as well as $I = 3/2$ $K + \pi$ states.

To summarise, for the $\kappa$ sector $I(J^P) = 1/2(0^+)$ mixing due to twisted mass symmetry breaking will double the number of two-particle $K + \pi$ states contained in our correlation matrices. This theoretical expectation is confirmed by our numerical results (cf.\ section~\ref{SEC448}).

% *****

\subsubsection{\label{SEC649}Different twisted mass realizations of the $s$ quark}

As mentioned in the previous section, we realize the $s$ quark via $s^+$.

Using $s^-$ would yield on a quantitative level slightly different numerical results. The reason is that one would observe a $(K +\pi) \equiv (s^- u + (\bar{u} u - \bar{d}) d)$ state and a $(K + \pi) \equiv (s^- d + \bar{d} u)$ state, i.e.\ each of the two states contains one ``heavy version'' of a meson and one ``light version'' of a meson. Of course, in the continuum limit $s^+$ and $s^-$ yield identical results.

The results presented in this paper exclusively correspond to $s^+$.

% **********
% **********
% **********

\subsection{Computation of correlation matrices}

We compute separate correlation matrices for $a_0(980)$ and $\kappa$,
\begin{eqnarray}
\label{EQN174} C_{j k}(t_2-t_1) \ \ = \ \ \Big\langle (\mathcal{O}_j(t_2))^\dagger \mathcal{O}_k(t_1) \Big\rangle ,
\end{eqnarray}
where $\mathcal{O}_j$ and $\mathcal{O}_k$ denote the creation operators (\ref{EQN002}) to (\ref{EQN523}). Technical details of these computations are explained in the following.

% *****

\subsubsection{Smearing techniques}

To improve the overlap to the low-lying states of interest, we use Gaussian smearing
of quark fields \cite{Gusken:1989qx,Alexandrou:2008tn} with APE smeared spatial
links \cite{Albanese:1987ds}. Detailed equations can be found e.g.\ in
\cite{Jansen:2008si}. We use APE smearing parameters $\alpha_\textrm{APE} = 0.5$ and $N_\textrm{APE} = 20$. Gaussian smearing is done with $\kappa_\textrm{Gauss} = 0.5$ and $N_\textrm{Gauss} = 50$ for most ensembles. Only for A40.20 we used $N_\textrm{Gauss} = 30$ instead of $N_\textrm{Gauss} = 50$. For lattice spacing $a \approx 0.086 \, \textrm{fm}$ these parameters are inside a region, in which the overlap between mesonic trial states and the $K$ and $D$ meson is rather large \cite{Baron:2010th}.

% *****

\subsubsection{\label{SEC229}Propagator computation}

For correlation matrix elements (\ref{EQN174}), where both $\mathcal{O}_j$ and $\mathcal{O}_k$ are molecule and/or diquark operators, 
we use point source inversions, i.e.\ twelve inversions per gauge link configuration and quark flavour. 
This yields point-to-all propagators, which are exact, but which do not allow to exploit spatial translational invariance of the correlation matrix elements,
to increase their statistical precision. In order to reduce correlations, however, we have chosen a random position for the source vector for each gauge configuration.

For correlation matrix elements, where at least one of the operators $\mathcal{O}_j$ and $\mathcal{O}_k$ is a two-particle operator, 
the situation is different: here the standard one-end trick \cite{Foster:1998vw} can be applied twice, allowing a stochastic estimation of timeslice-to-all propagators. 
For each application of the one-end-trick we generated an independent stochastic timeslice source with $\mathcal{Z}_2 \times \mathcal{Z}_2$ noise, 
where the source time slice has been chosen randomly for each gauge configuration.
Computing correlation matrix elements with stochastic timeslice-to-all propagators is rather efficient, because they allow to exploit spatial translational invariance, 
which in turn reduces gauge noise significantly. 
Moreover, correlations between two two-particle operators require timeslice-to-all propagators, 
which are prohibitively expensive to compute using point source inversions.

% **********
% **********
% **********

\subsection{Analysis of correlation matrices}

To extract energy levels from $N \times N$ correlation matrices, we solve the generalised eigenvalue problem
\begin{eqnarray}
C(t) \vec{v}^n(t) \ \ = \ \ \lambda^n(t,t_0) C(t_0) \vec{v}^n(t) \quad , \quad n = 0,\ldots,N-1
\end{eqnarray}
(cf.\ e.g.\ \cite{Blossier:2009kd} and references therein). For a lattice with infinite temporal extension $T$ the eigenvalues $\lambda^n(t,t_0)$ are proportional to $e^{-E_n t}$ for sufficiently large $t$, where $E_n$ are the energies of the $N$ lowest energy eigenstates.

However, for lattices with periodic finite temporal extensions and
sectors, where light two-particle states exist (in our case two
pseudoscalar mesons; cf.\ sections \ref{SEC483} and \ref{SEC484}), the
interpretation of the eigenvalues $\lambda^n(t,t_0)$ is no longer
simple. For example a diagonal correlator $C_{j j}(t)$, which is
dominated by a two-particle state with energy $E_n$ shows the
following behaviour for $0 \ll t \ll T$ \cite{Prelovsek:2010kg,
  Detmold:2008yn, Prelovsek:2008rf}: 
\begin{eqnarray}
\label{eqn:single-particle} C_{j j}(t) \ \ = \ \ \Big|A_j^n\Big|^2 \bigg(e^{-E_n t} + e^{-E_n (T-t)}\bigg) + \Big|B_j^n\Big|^2 \bigg(e^{-m_1 t} e^{-m_2 (T-t)} + e^{-m_2 t} e^{-m_1 (T-t)}\bigg) ,
\end{eqnarray}
where $m_1$ and $m_2$ denote the masses of the corresponding single-particle states and $A_j^n$ and $B_j^n$ are operator dependent and problem specific constants. The ``$B_j^n$ term'' corresponds to the ``$m_1$ particle'' traveling forward in time, while the ``$m_2$ particle'' is traveling backwards in time, and vice versa. This term leads to a drastic and characteristic deviation of effective masses from their plateaux values at larger temporal separations. For example in Figure~\ref{FIG001}b this effect is clearly visible for $t/a \gtapprox 15$.

In \cite{Prelovsek:2010kg} eq.\ (\ref{eqn:single-particle}) was fitted to the eigenvalues $\lambda^n(t,t_0)$ to extract the energy levels $E_n$. In this project, however, several two-particle states with rather different single-particle masses contribute: in the $a_0(980)$ sector $K + \bar{K}$ and $\eta_s + \pi$ are the relevant states (cf.\ section~\ref{SEC483}); for $\kappa$, due to twisted mass symmetry breaking, light and heavy kaons and pions need to be considered (cf.\ section~\ref{SEC484}). Since a proper treatment of all these two-particle states yields an equation significantly more complicated than (\ref{eqn:single-particle}) with too many parameters to perform stable fits, we follow a different route.

First note that the $B_j^n$ term in (\ref{eqn:single-particle}) is
suppressed by $\approx e^{-\min(m_1,m_2) T}$. Since $A_j^n$ and
$B_j^n$ are of the same order of magnitude, the $B_j^n$ term becomes
irrelevant for sufficiently small $t$ or $T-t$. Hence, we extract the
energy levels considering small temporal separations only. We restrict
all our effective mass analyses to $t , T-t \ltapprox T/4$, which
seems to be a rather conservative choice. Possibly present excited
state contributions are taken into account by fitting two exponentials
to each of the eigenvalues $\lambda^n(t,t_0)$ of interest, fitting
range $t_\textrm{min} \leq t \leq t_\textrm{max}$. $t_0$,
$t_\textrm{min}$ and $t_\textrm{max}$ have been chosen such that
$\chi^2/\text{dof} \ltapprox 1$. Moreover, we varied the values of
$t_0$, $t_\textrm{min}$ and $t_\textrm{max}$ to check and confirm the
stability of our fitting results.

% ********************
% ********************
% ********************
% ********************
% ********************
\section{\label{SEC448}Numerical results and their interpretation}

% **********
% **********
% **********

\subsection{\label{SEC129}$a_0(980)$: tetraquark and two-particle operators, ensemble A40.20}

We start by discussing $a_0(980)$ ($I(J^P) = 1/2 (0^+)$) results obtained using ensemble A40.20 (cf.\ Table~\ref{TAB001}). This ensemble with rather small spatial extension ($L \approx 1.72 \, \textrm{fm}$) is particularly suited to distinguish two-particle states with relative momentum from states with two particles at rest and from possibly existing $a_0(980)$ tetraquark states (two-particle states with relative momentum have a rather large energy because one quantum of momentum $p_\textrm{min} = 2 \pi / L \approx 720 \, \textrm{MeV}$).

\begin{figure}[htbp]
\begin{center}
 \begin{tabular}{ccc}
 (a) & & (b) \\
 \includegraphics[width=0.45\textwidth]{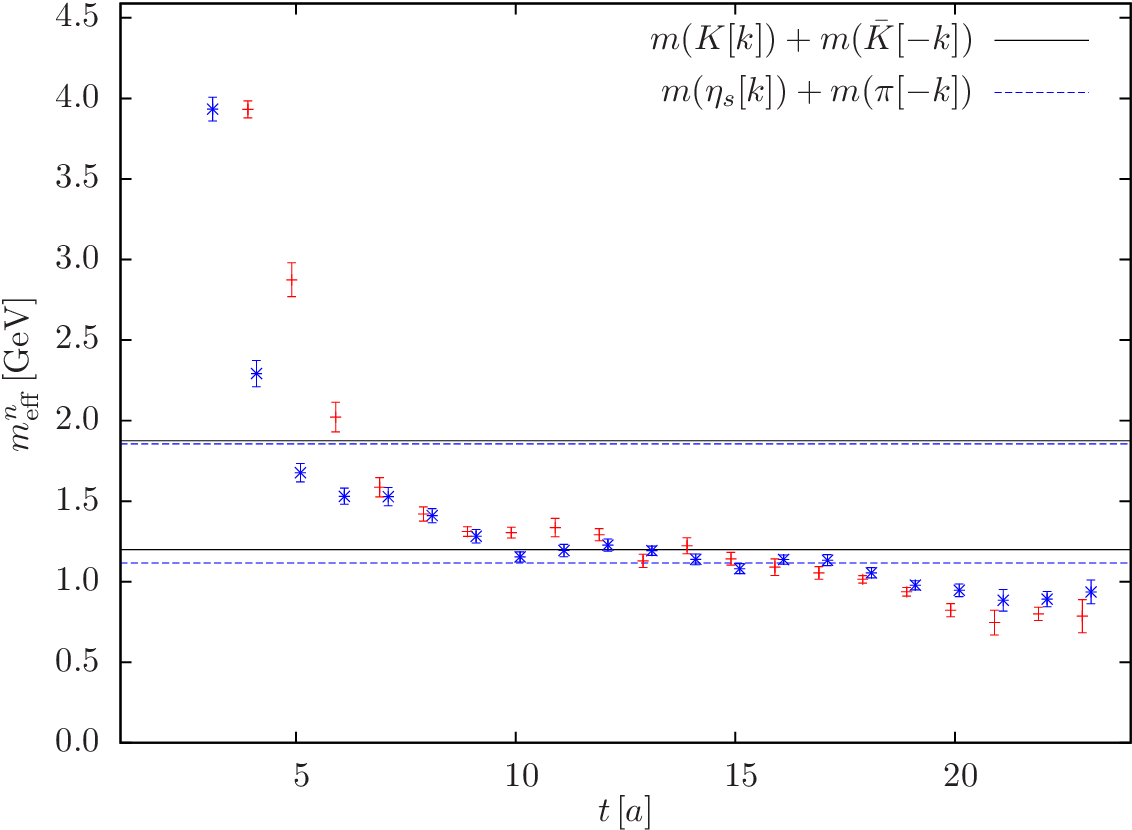} & & 
 \includegraphics[width=0.45\textwidth]{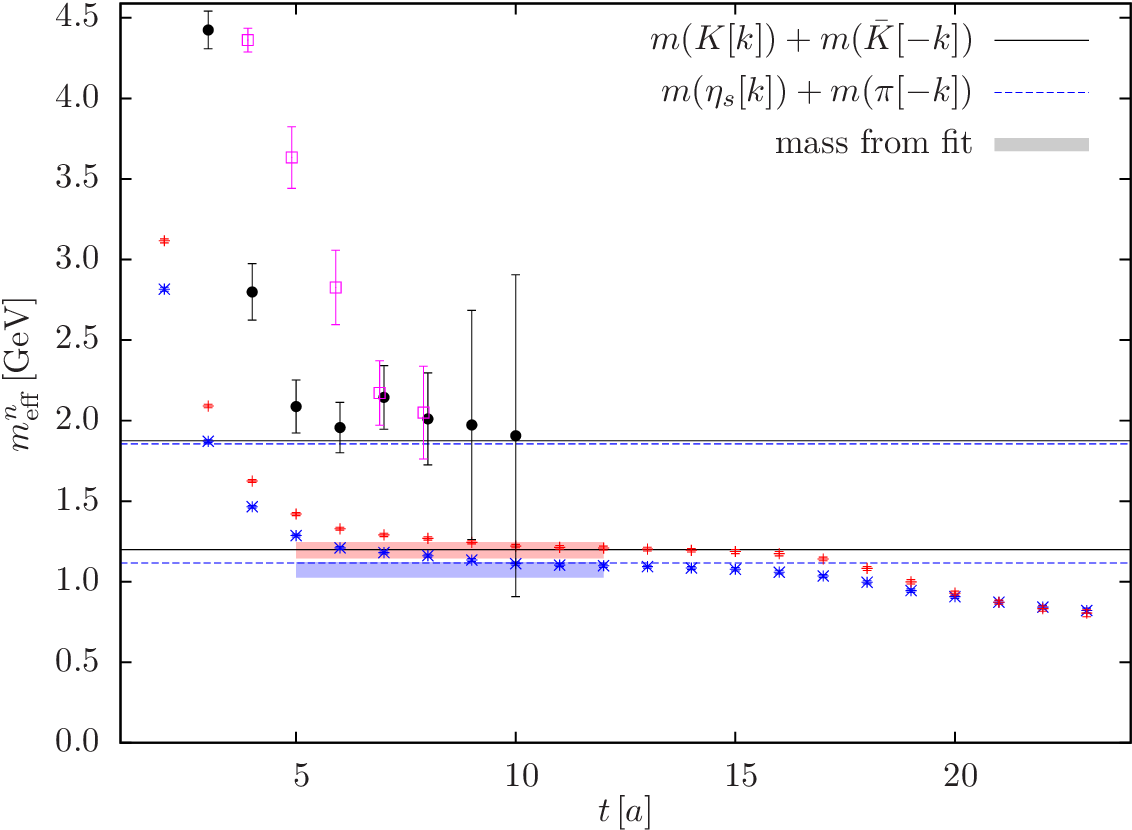} \\   & & \\
 (c) &  & (d) \\
 \includegraphics[width=0.45\textwidth]{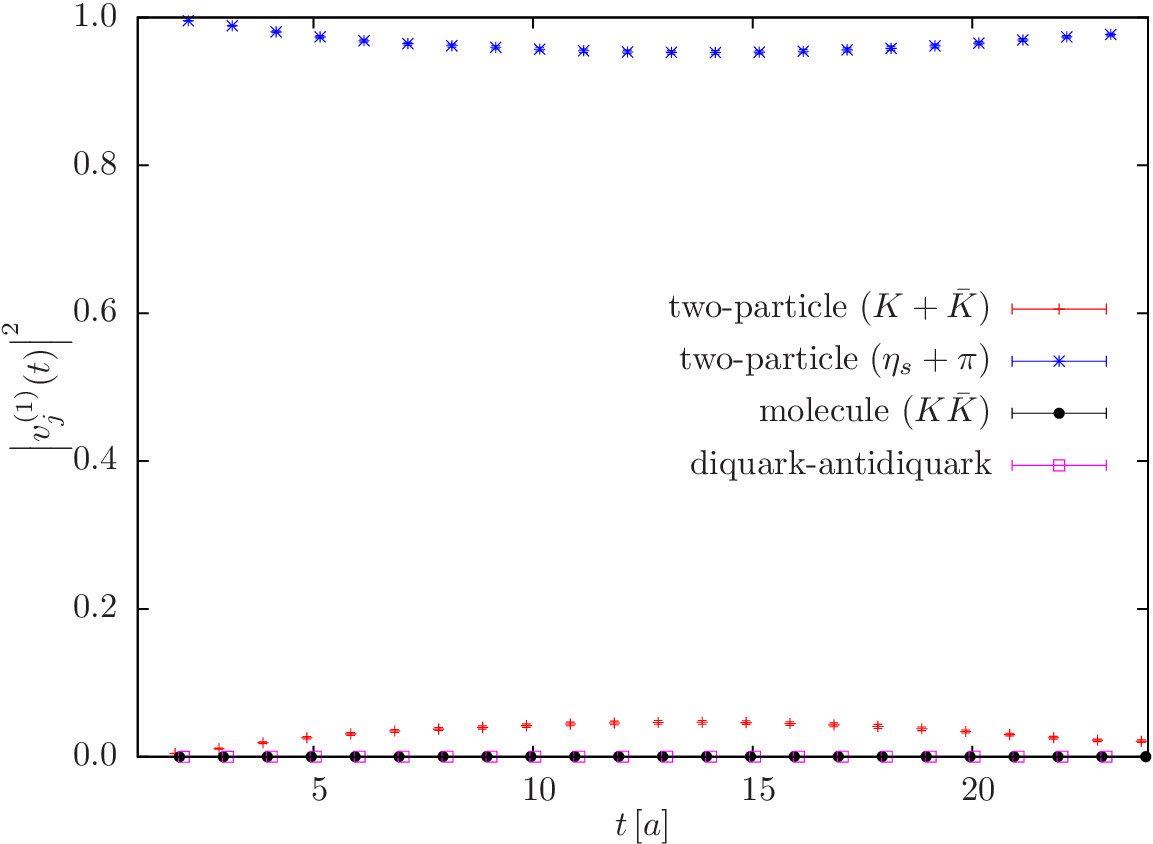}  & &
 \includegraphics[width=0.45\textwidth]{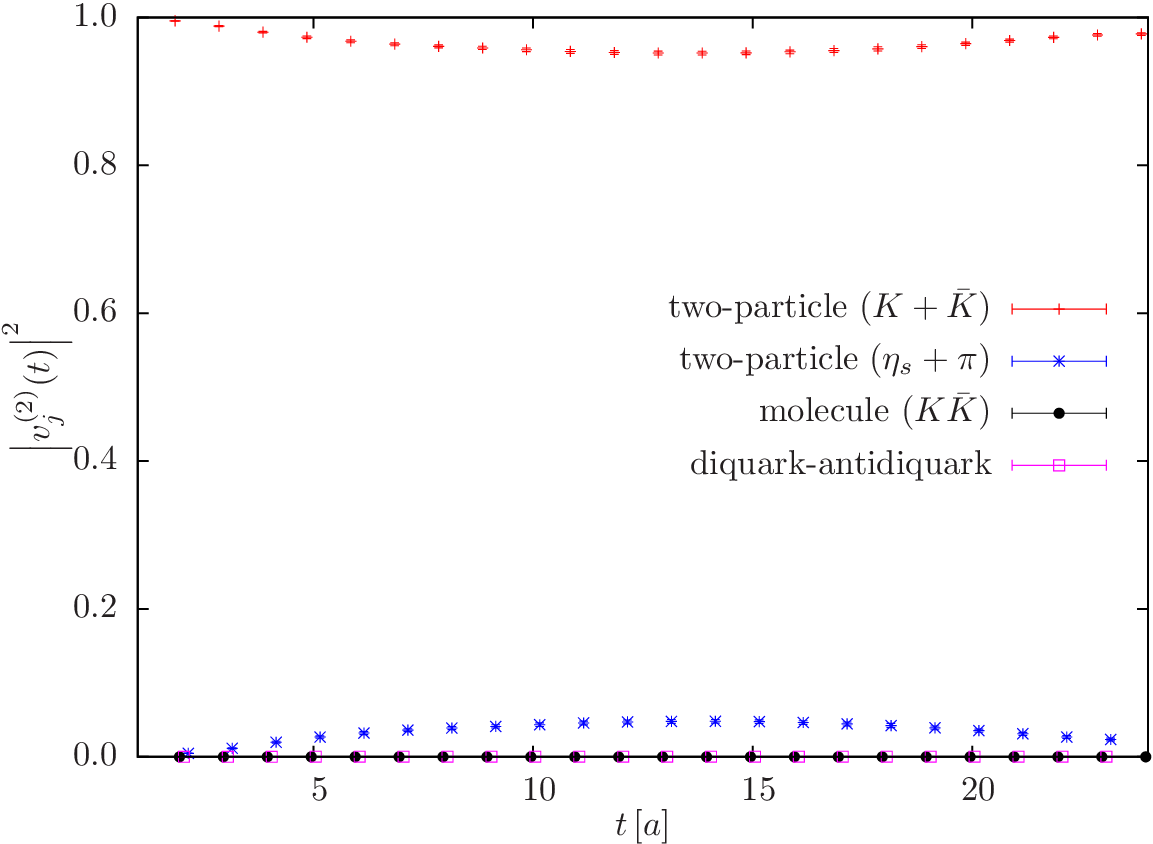} \\
 \end{tabular}

\end{center}

\caption{\label{FIG001}$a_0(980)$ sector, A40.20 ensemble.
\textbf{(a)}~Effective masses as a function of the temporal
separation, $2 \times 2$ correlation matrix (local operators: $K
\bar{K}$ molecule, diquark-antidiquark, eqs.\ (\ref{EQN002}) and
(\ref{EQN524})). Horizontal lines indicate the expected two-particle
$K + \bar{K}$ and $\eta_s + \pi$ energy levels. 
\textbf{(b)}~$4 \times 4$ correlation matrix (local operators: $K \bar{K}$ molecule, diquark-antidiquark, two-particle $K + \bar{K}$, two-particle $\eta_s + \pi$, eqs.\ (\ref{EQN002}) to (\ref{EQN003})).
\textbf{(c)}, \textbf{(d)}~Squared eigenvector components of the two
low-lying states from \textbf{(b)} as a function of the temporal separation.}

\end{figure}

Figure~\ref{FIG001}a shows effective mass plots from a $2 \times 2$
correlation matrix with a $K \bar{K}$ molecule operator (\ref{EQN002})
and a diquark-antidiquark operator (\ref{EQN524}), flavour combination
$s^+ \bar{s}^-$ (cf.\ section~\ref{SEC648}). The corresponding
energies extracted from the two plateaus are given in
Table~\ref{TAB599} and they are consistent both with the expectation
for possibly existing $a_0(980)$ tetraquark states and with
two-particle $K + \bar{K}$ and $\eta_s + \pi$ states, where both
particles are at rest ($m(K + \bar{K}) \approx 2 m(K) \approx 1198 \,
\textrm{MeV}$; $m(\eta_s + \pi) \approx m(\eta_s) + m(\pi) \approx
1115 \, \textrm{MeV}$; cf.\ Table~\ref{TAB011}). 

Increasing this correlation matrix to $4 \times 4$ by adding
two-particle $K + \bar{K}$ and $\eta_s + \pi$ operators (eqs.\
(\ref{EQN093}) and (\ref{EQN003})) yields the effective mass results
shown in Figure~\ref{FIG001}b. Two additional states are observed with
energies given in Table~\ref{TAB599}. From this $4 \times 4$ analysis
we conclude the following:
\begin{itemize}
\item We do not observe a third low-lying state around $1000 \,
  \textrm{MeV}$, even though we provide operators, which are of
  tetraquark type as well as of two-particle type. This suggests that
  the two low-lying states are the expected two-particle $K + \bar{K}$
  and $\eta_s + \pi$ states, while no additional stable $a_0(980)$
  tetraquark state is detected.

\item The effective masses of the two low-lying states are of much
  better quality in Figure~\ref{FIG001}b than in
  Figure~\ref{FIG001}a. We attribute this to the two-particle $K +
  \bar{K}$ and $\eta_s + \pi$ operators, which appear to create larger
  overlap to those states than the tetraquark operators. This in turn
  confirms the interpretation of the two low-lying states as
  two-particle states. 

\item To investigate the overlap in a more quantitative way, we show the squared eigenvector components of the two low-lying states in Figure~\ref{FIG001}c and Figure~\ref{FIG001}d (cf.\ \cite{Baron:2010vp} for a more detailed discussion of such eigenvector components). Clearly, the lowest state is of $\eta_s + \pi$ type, whereas the second lowest state is of $K + \bar{K}$ type. On the other hand, the two tetraquark operators are essentially irrelevant for resolving those states, i.e.\ they do not seem to contribute any structure, which is not already present in the two-particle operators. These eigenvector plots give additional strong support of the above interpretation of the two low lying states as two-particle states.

\item The energy of two-particle excitations with one relative quantum of momentum can be estimated by
\begin{eqnarray}
\label{EQN775} m(1 + 2,p = p_\textrm{min}) \ \ \approx \ \ \sqrt{m(1)^2 + p_\textrm{min}^2} + \sqrt{m(2)^2 + p_\textrm{min}^2} \quad , \quad p_\textrm{min} \ \ = \ \ \frac{2 \pi}{L} .
\end{eqnarray}
Inserting $m(K)$, $m(\eta_s)$ and $m(\pi)$ from Table~\ref{TAB011}, yields $m(K + \bar{K},p = p_\textrm{min}) \approx 1873 \, \textrm{MeV}$ and $m(\eta_s + \pi,p = p_\textrm{min}) \approx 1853 \, \textrm{MeV}$ for the A40.20 ensemble. These numbers are consistent with the effective mass plateaus of the second and third excitation in Figure~\ref{FIG001}b. Consequently, we also interpret them as two-particle states.
\end{itemize}

Figure~\ref{FIG001}a and Figure~\ref{FIG001}b also demonstrate an important technical aspect: two-particle states can be resolved by tetraquark operators, i.e.\ two-particle operators are not necessarily needed, to extract the full spectrum. Since we are mainly interested in possibly existing states with a strong tetraquark component, we restrict the correlation matrices computed for other ensembles to four-quark operators (cf.\ Table~\ref{TAB569}). This allows to save a significant amount of computer time, because two-particle operators require different inversions of the Dirac operator (timeslice propagators instead of point propagators; cf.\ section~\ref{SEC229}).

% **********
% **********
% **********

\subsection{$a_0(980)$: tetraquark operators, many ensembles}

We have analysed the six ensembles listed in Table~\ref{TAB001} with respect to $a_0(980)$ in a similar way as explained in section~\ref{SEC129}.

As already mentioned above the main difference is that this time we exclusively use tetraquark operators (\ref{EQN002}) to (\ref{EQN524}), but no two-particle operators (\ref{EQN093}) and (\ref{EQN003}). To be able to resolve more than two low-lying states, we supplement the molecule operators and the diquark-antidiquark operator by versions, where $\gamma_5$ has been replaced by $\gamma_j$ and $\gamma_j \gamma_5$ (molecule) and by $1$ (diquark-antidiquark). More detailed information including e.g.\ smearing parameters, number of gauge link configurations, etc.\ are collected in Table~\ref{TAB569}.

\begin{figure}[p]
\begin{center}
\begin{tabular}{ccc}
 A30.32 (s$^+\bar{s}^+$) & &  A40.20 (s$^+\bar{s}^+$) \\
\includegraphics[width=0.45\textwidth]{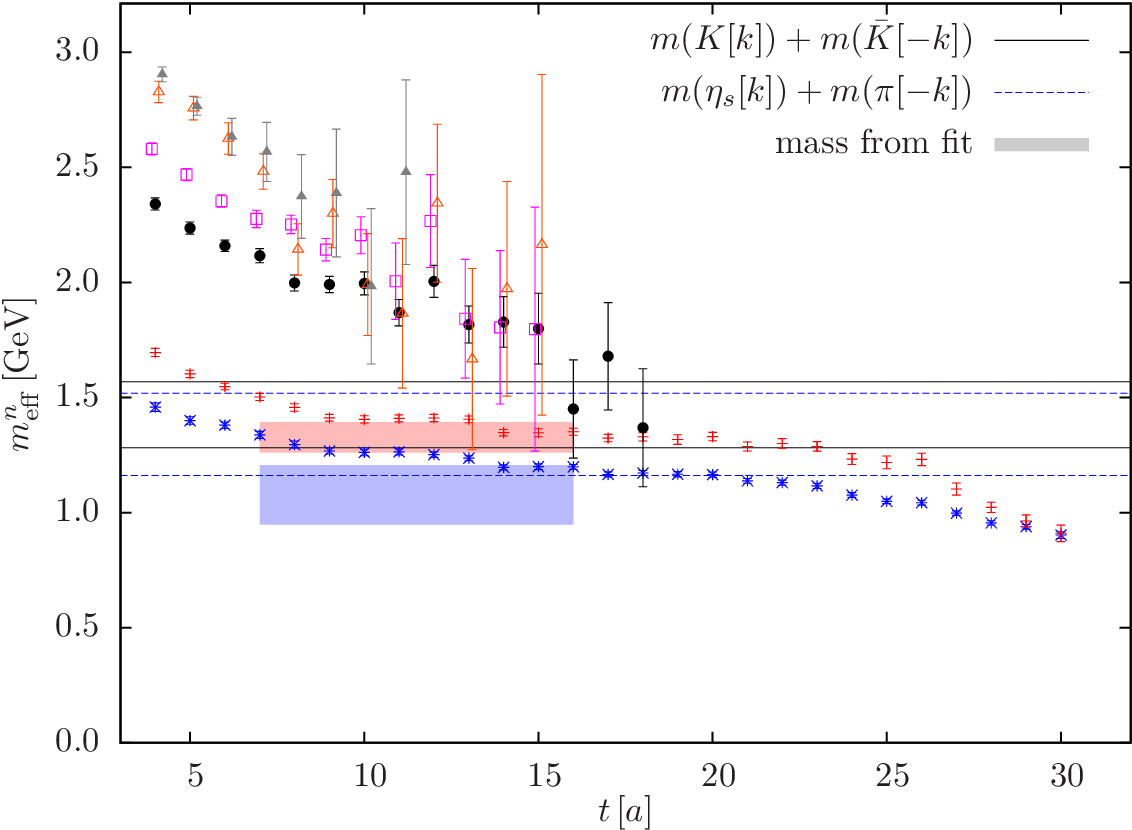} & &
\includegraphics[width=0.45\textwidth]{A40.20a0.ps}  \\ & &  \\
 A40.24 (s$^+\bar{s}^+$) & & A40.32 (s$^+\bar{s}^+$) \\
\includegraphics[width=0.45\textwidth]{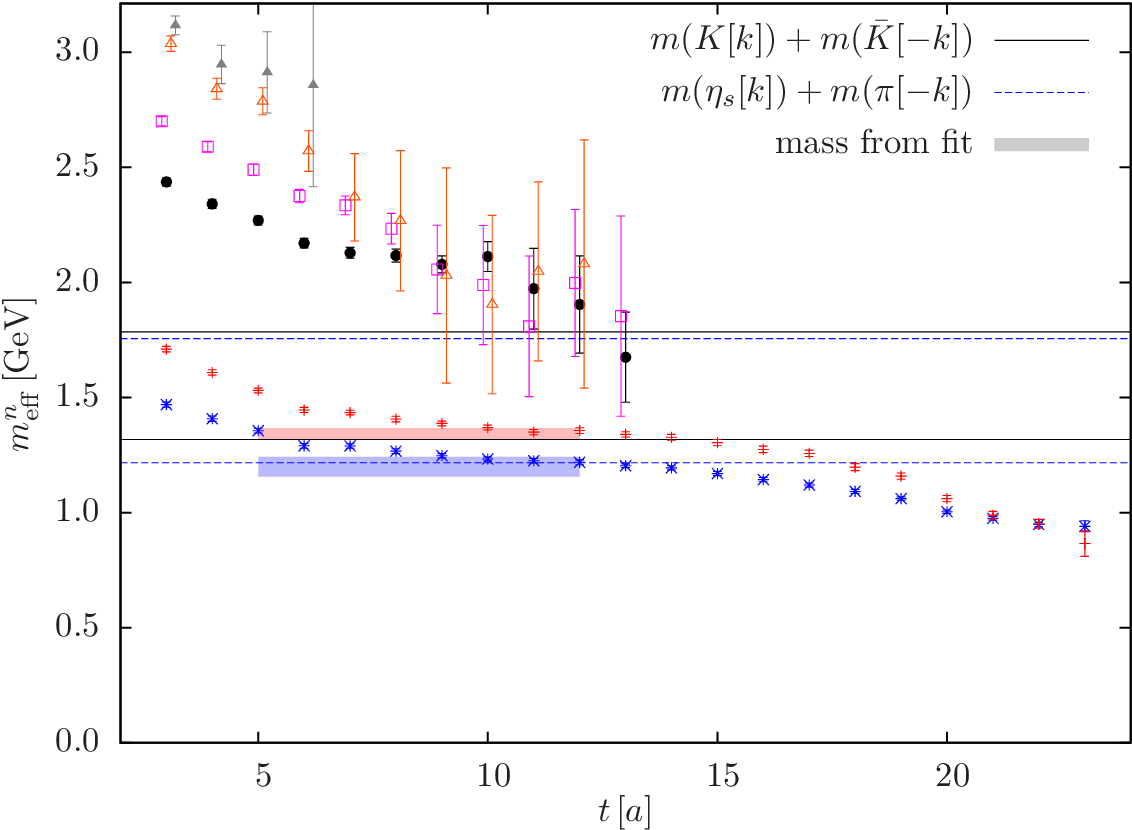} & &
\includegraphics[width=0.45\textwidth]{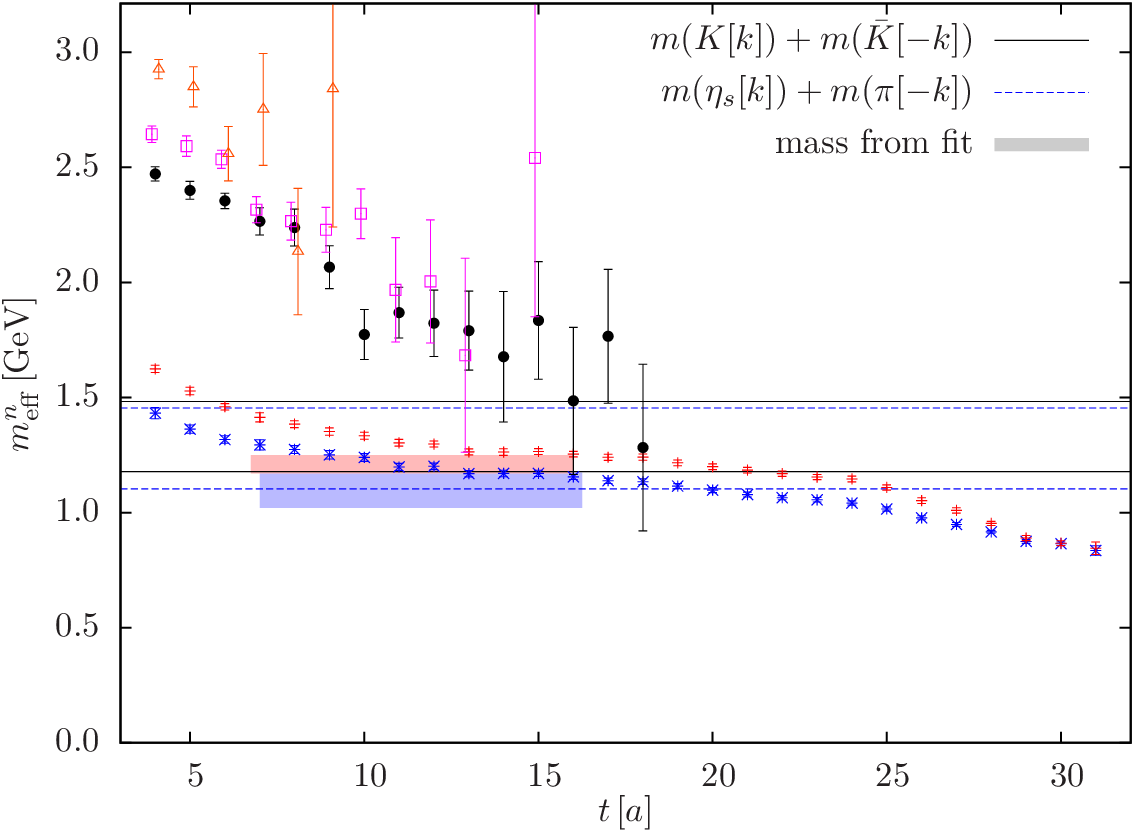}  \\ & & \\
 A80.24 (s$^+\bar{s}^+$) & & A50.32 (s$^+\bar{s}^+$) \\
\includegraphics[width=0.45\textwidth]{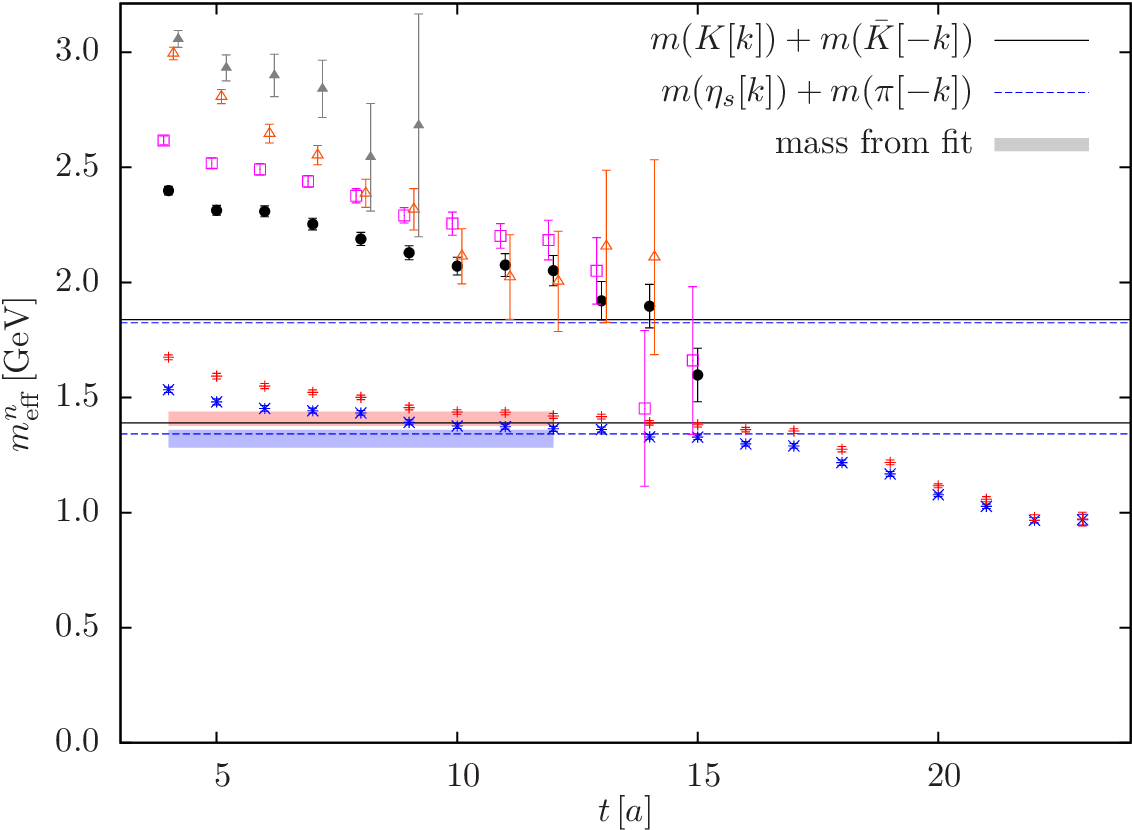} & &
\includegraphics[width=0.45\textwidth]{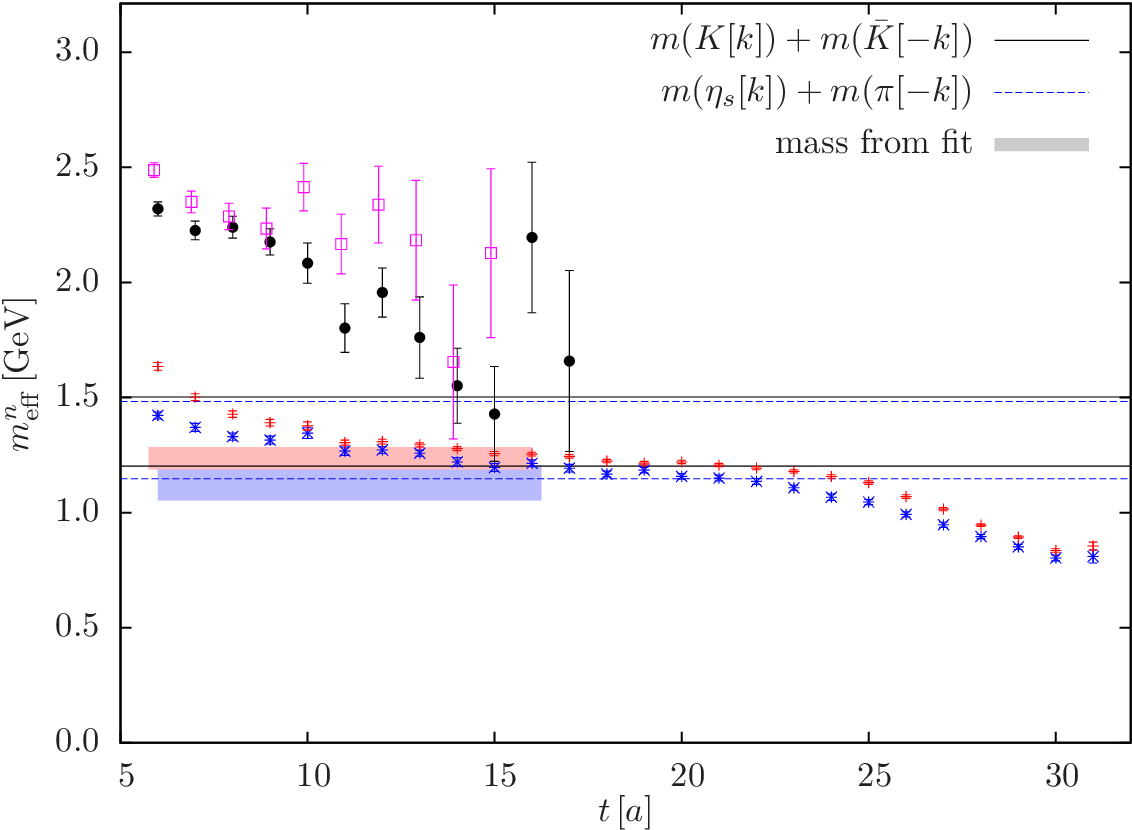}  \\ &  & \\
\end{tabular}

\caption{\label{FIG005}$a_0(980)$ sector, various ensembles, some of
  them with twisted mass strange quarks $s^+ \bar{s}^+$, others with
  $s^+ \bar{s}^-$. Effective masses as a function of the temporal
  separation. Horizontal lines indicate the expected two-particle $K +
  \bar{K}$ and $\eta_s + \pi$ energy levels.} 
\end{center}
\end{figure}

\begin{figure}[htbp]
\begin{center}

\begin{tabular}{ccc}
 $a_0(980)$ sector & & $\kappa$ sector  \\
\includegraphics[width=0.45\textwidth]{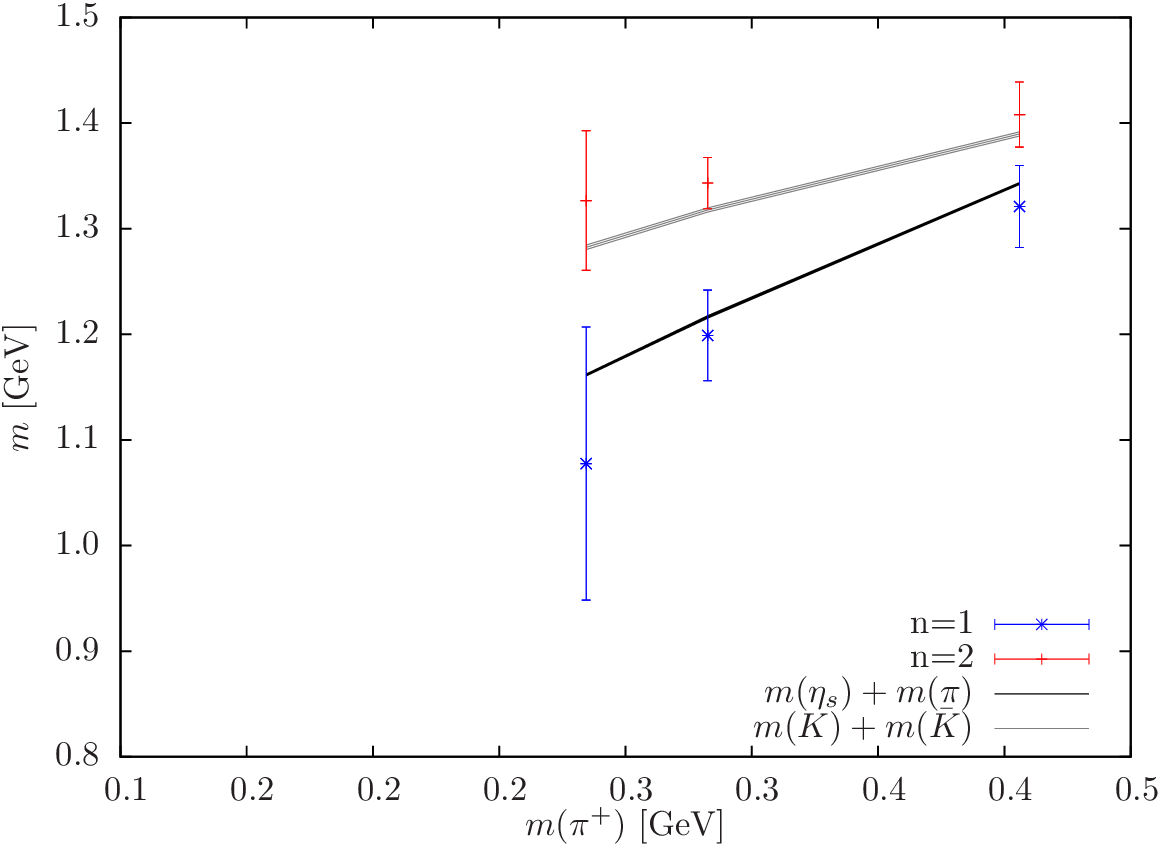} & &
\includegraphics[width=0.45\textwidth]{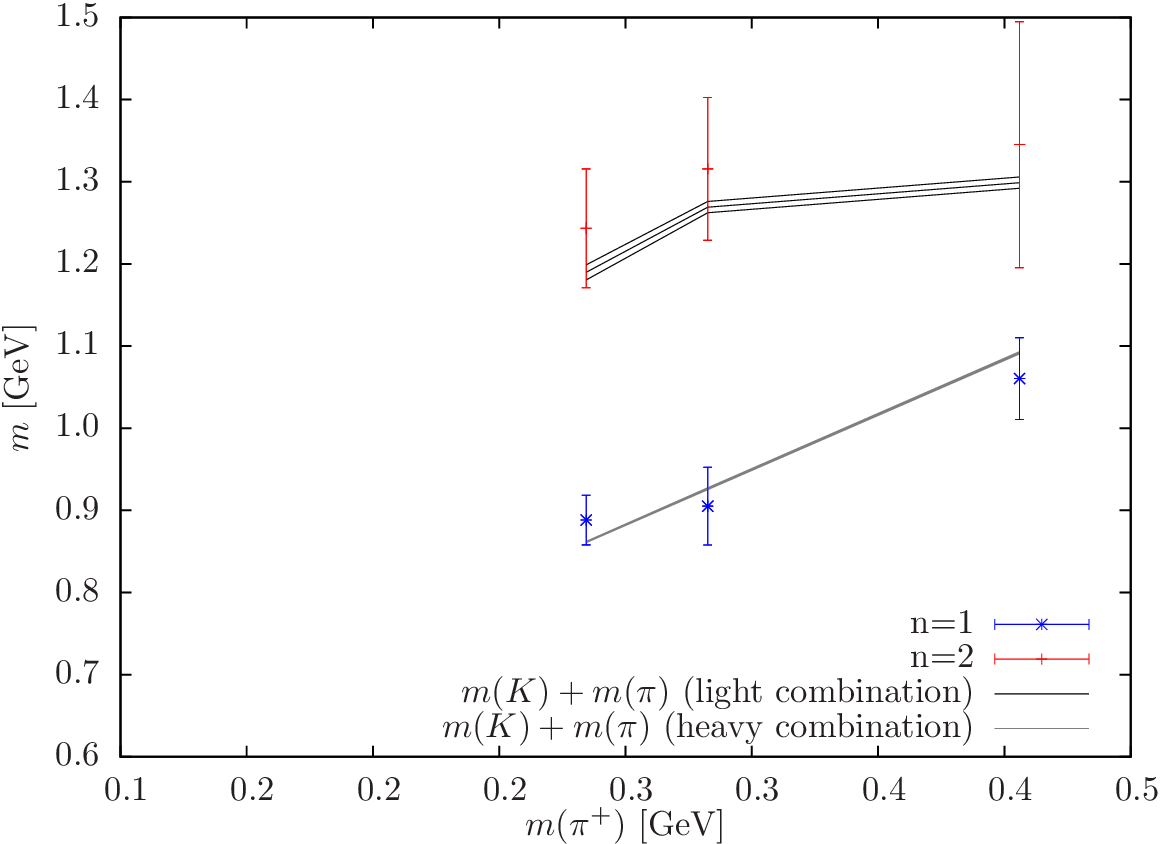} \\
\end{tabular}
\caption{\label{FIG752}The two lowest energy levels $E_0$ and $E_1$ obtained by our simulations in the $a_0(980)$ ($s^+\bar{s}^+$ only) and in the $\kappa$ sector (cf.\ also Table~\ref{TAB599}) as a function of $m_{\pi^+}$. Additionally, the energy levels of the non-interacting states are included as solid lines.}
\end{center}
\end{figure}

On a qualitative level our findings agree for all ensembles, i.e.\ are as reported in the previous subsection (effective mass plots are collected in Figure~\ref{FIG005}): there are always two low-lying states, whose masses are consistent with the expected masses of the two-particle $K + \bar{K}$ and $\eta_s + \pi$ states (cf.\ Figure~\ref{FIG752} and Table~\ref{TAB599}); higher excitations (the third, forth, etc.\ extracted state) are in all cases significantly heavier and consistent with two-particle excitations with one relative quantum of momentum (cf.\ eq.\ (\ref{EQN775})).

\begin{table}[htb]
\centering
\begin{tabular}{ccccccc}
\toprule
 & ensemble & $t_0$ & $t_\text{min}$ & $t_\text{max}$ & $E_0$ in MeV & $E_1$ in MeV \\
\midrule
$a_0(980)$    &      A40.24    &  2  &  5         &  
12         & 1199(43)\phantom{0}   &           1343(24)\phantom{0}        \\
    ($s^+ \bar{s}^+$)                                  &      A30.32    &  3  &   
7         & 16         & 1078(129)  &           1327(66)\phantom{0}        \\
                                      &      A80.24    &  3  &  4         &  
12         & 1321(39)\phantom{0}   &           1408(31)\phantom{0}        \\
\hline
$a_0(980)$   &      A40.20    &  1  &  5         & 12          
& 1073(48)\phantom{0}   &            1195(51)\phantom{0}        \\
   ($s^+ \bar{s}^-$)                                    &      A40.32    &  3  &  
  7         & 16         & 1098(77)\phantom{0}   &           1210(40)\phantom{0}        \\
                                      &      A50.32    &  5  &  6         &  
16         & 1130(77)\phantom{0}   &           1236(48)\phantom{0}        \\
\hline
     $\kappa$                                     &       
A30.32 &  3  &  7         & 16         & \phantom{0}888(30)\phantom{0}    &           1243(72)\phantom{0}        \\
                                      &      A40.24 &  3  &  5         & 12          
& \phantom{0}905(47)\phantom{0}    &           1316(87)\phantom{0}        \\
                                      &      A80.24 &  3  &  5         & 12         & 1060(50)\phantom{0}   &           1345(150)       \\

\bottomrule
\end{tabular}
\caption{\label{TAB599}The two lowest energy levels $E_0$ and $E_1$ in the $a_0(980)$ and in the $\kappa$ sector (cf.\ also Figure~\ref{FIG752}).}
\end{table}

To summarise, in the lattice setup and ensembles we are studying there is no indication of any additional low-lying tetraquark state.

% **********
% **********
% **********

\subsection{\label{SEC001}$\kappa$: tetraquark operators, many ensembles}

The analysis for the $\kappa$ sector ($I(J^P) = 1/2 (0^+)$) closely parallels the analysis of the $a_0(980)$ sector presented above.

We consider correlation matrices containing a $K \pi$ molecule operator (\ref{EQN522}) and analogue versions with $\gamma_5$ replaced by $\gamma_j$ and $\gamma_j \gamma_5$ as well as an diquark-antidiquark operator ((\ref{EQN523}) and a similar operator with $\gamma_5$ replaced by $1$). More detailed information including e.g.\ smearing parameters, number of gauge link configurations, etc.\ are collected in Table~\ref{TAB569}.

As has been explained in section~\ref{SEC329} in twisted mass lattice QCD isospin $I$ is not a quantum number. Therefore, it is not sufficient to only resolve $I = 1/2$ two-particle $K + \pi$ states. One has to take into account also mixing with $I = 3/2$ two-particle $K + \pi$ states, i.e.\ it is necessary to resolve these two types of low-lying two-particle states at the same time.

\begin{figure}[htbp]
\begin{center}

\begin{tabular}{ccc}
 A30.32  & &  A40.24 \\
\includegraphics[width=0.45\textwidth]{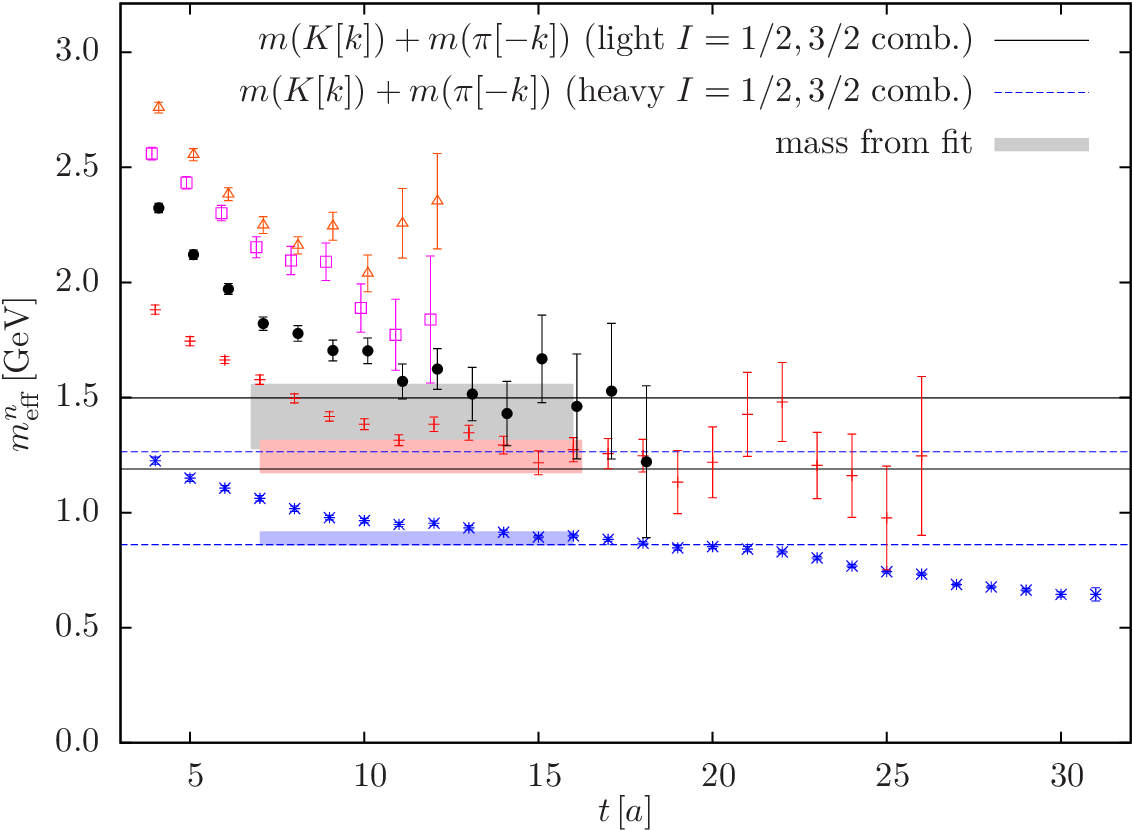} & &
\includegraphics[width=0.45\textwidth]{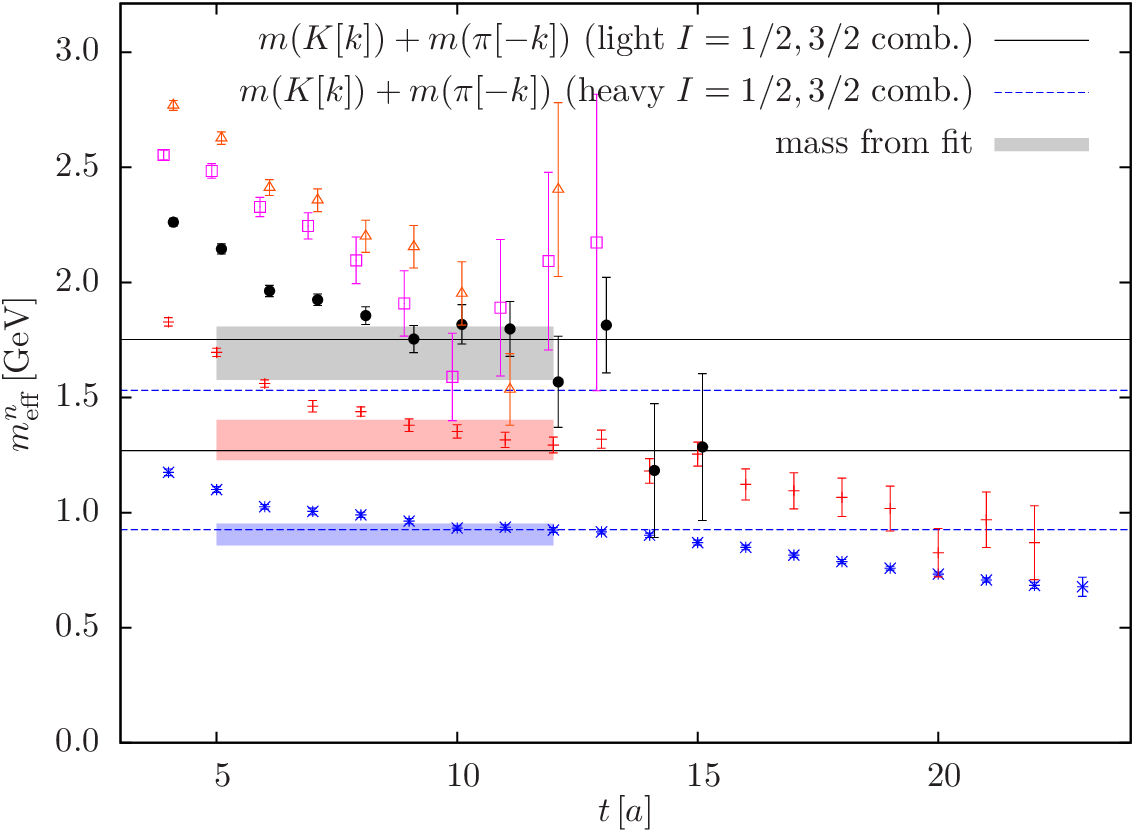}  \\ & &  \\
 A80.24  & &  \\
\includegraphics[width=0.45\textwidth]{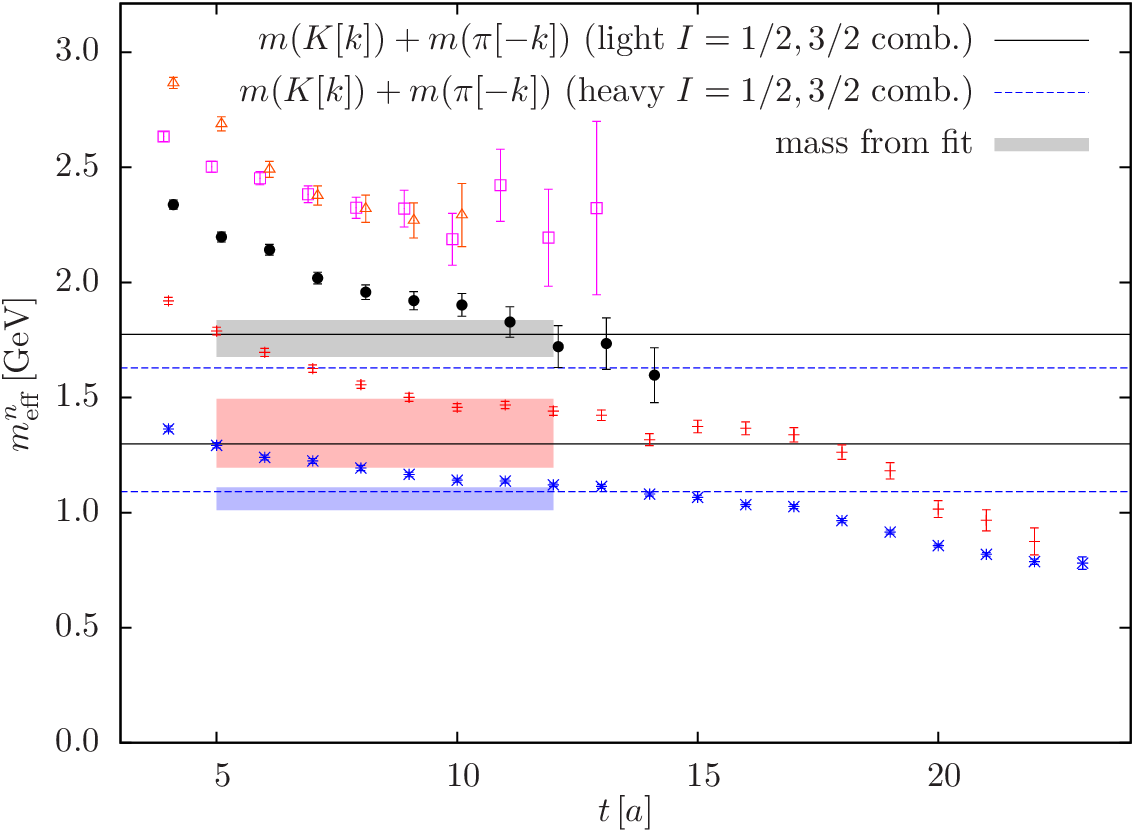} & & \\ 
\end{tabular}

\caption{\label{FIG003}$\kappa$ sector, various ensembles. Effective
  masses as a function of the temporal separation. Horizontal lines
  indicate the expected two-particle $K + \pi$ energy levels.} 
\end{center}
\end{figure}

Effective mass plots for ensembles A30.32, A40.24 and A80.24 (cf.\ Table~\ref{TAB001}) are shown in Figure~\ref{FIG003} together with the expected energy levels of two-particle $K + \pi$ states (obtained via eq.\ (\ref{EQN775}) and the meson masses collected in Table~\ref{TAB011}). While effective mass plateaus are consistent with these expected two-particle energy levels, there is no indication of any additional low lying state, i.e.\ of a possibly existing bound $\kappa$ state. While this is suggested by experimental data, it contradicts the findings of a similar recent lattice study of $\kappa$ \cite{Prelovsek:2010kg}. Currently we have no explanation for this qualitative discrepancy of two rather similar lattice computations (same operators, no disconnected diagrams, similar quark masses).

Results for the two lowest energy levels are collected in Figure~\ref{FIG752} and Table~\ref{TAB599}.

% ********************
% ********************
% ********************
% ********************
% ********************

\section{Conclusions and outlook}

This work represents a first necessary step in the long term project
of studying the scalar mesons and their properties 
on the lattice.  The main goal of this work was to develop and test
those techniques that can be effectively 
exploited for studying the contribution of four-quark operators in
mesons, especially in the context of the twisted 
mass formulation of lattice QCD.

In particular we computed the low-lying spectrum in the $a_0(980)$ and
$\kappa$ sectors by employing trial states designed to have a
substantial overlap with both two-particle and possibly existing
tetraquark states. With our ensembles we 
did not see additional states beside those that can be identified with
the expected two-particle spectrum. In fact for all our ensembles we
observed two low lying states in correspondence with the $K+\bar{K}$
and $\eta_s + \pi$ 
thresholds in the $a_0(980)$ sector and the $K+\pi$ ($I = 1/2$ and
$I = 3/2$) threshold in the $\kappa$ sector.  The next states appear
roughly consistent with excitations of the first quantum of momentum 
($2\pi/L$) on top of those thresholds.  This is somewhat difficult to
reconcile with the additional state found by 
\cite{Prelovsek:2010kg} in the $\kappa$, in spite of the rather
similar lattice setups. 

We find that the low lying spectrum has essentially exclusively
overlap to two-particle trial states.  This 
suggests that the states that we see are, indeed, the expected
two-particles states at the threshold and not 
tightly bound states either of molecular type or diquark-antidiquark
type. 

On the basis of this, we can conclude that either our choice of
operators has negligible overlap with the wave 
function of the resonances $a_0(980)$ and $\kappa$, or that our
volumes are not large enough to identify those 
states. 

These conclusions can be strengthened by studying more volumes, by
introducing twisted boundary 
conditions \cite{Bedaque:2004kc} and by studying further trial states
of different type.  As for the latter, it will be crucial to 
combine four quarks with traditional quark-antiquark operators, but
disconnected diagrams will be necessary for that. 
As for the volume dependence, we plan to use the finite volume
formulae of L\"uscher \cite{Luscher:1985dn, 
Luscher:1986pf, Luscher:1990ck, Luscher:1990ux, Luscher:1991cf} and
their extensions to multiple channels developed 
in \cite{Lage:2009zv,Bernard:2010fp,Oset:2011ce,Doring:2011vk}.  At
present, our limited number of volumes is 
insufficient for such an analysis. Corresponding computations are in
progress. 

Another possible development consists in studying four-quark states
that include the charm quark.  This is a natural 
extension thanks also to the presence of a dynamical charm quark in
the ETMC gauge configurations.  This direction 
is also being explored in particular in the context of the tetraquark
candidates $D_{s0}^\ast$ and $D_{s1}^\ast$.

% ********************
% ********************
% ********************
% ********************
% ********************

\section*{Acknowledgements}

It is a great pleasure to thank Akaki Rusetsky for many enlightening
discussions. We also acknowledge helpful discussions with Vladimir
Galkin, Vincent Mathieu, Francesco Di Renzo and Christian Wiese. We
thank Konstantin Ottnad for providing analysis code, and Marco
Cristoforetti for his contribution in a preliminary phase of this
project.

M.G.\ was supported by the Marie-Curie European training network ITN
STRONGnet grant PITN-GA-2009-238353. L.S.\ and acknowledge
support from the AuroraScience project funded by the Province of
Trento and INFN. M.D.B. is currently funded by the Irish Research
Council, acknowledges support by STRONGnet and the AuroraScience project,
and is grateful for the hospitality at ECT* and the University of
Cyprus, where part of this work was carried out.
M.W.\ acknowledges support by the Emmy Noether Programme of the DFG
(German Research Foundation), grant WA 3000/1-1. This work is
supported in part by the DFG and the NSFC through funds provided to
the Sino-German CRC 110 ``Symmetries and the Emergence of Structure in
QCD''. For parts of this project we used computer time that was made
available to us by the John von Neumann-Institute for Computing (NIC)
on the JUDGE system in J{\"u}lich. In particular we thank
U.-G.~Mei{\ss}ner for granting us access on JUDGE.
This work was partly supported by funding received  from the
Cyprus Research Promotion Foundation under contracts PENEK/0609/17 and
KY-$\Gamma$A/0310/02 and the infrastructure project Cy-Tera  co-funded
by the European Regional Development Fund and the Republic of Cyprus
through the Research Promotion Foundation (Project Cy-Tera NEA
Y$\Pi$O$\Delta$OMH/$\Sigma$TPATH/0308/31). This work was supported in
part by the Helmholtz International Center for FAIR within the
framework of the LOEWE program launched by the state of Hesse. J.D\
and C.U.\ were supported by the Bonn-Cologne Graduate School (BCGS) of
Physics and Astronomie.

Part of the computations presented here were performed on the Aurora
system in Trento.

% ********************
% ********************
% ********************
% ********************
% ********************

% ********************

\end{document}